\documentclass[12pt,preprint]{aastex}

\newcommand{\bdv}[1]{\mbox{\boldmath$#1$}}

\def\au{{\rm AU}} 
 
\def\kms{{\rm km}\,{\rm s}^{-1}}
\def\masyr{{\rm mas}\,{\rm yr}^{-1}}
\def\kpc{{\rm kpc}}
\def\mas{{\rm mas}}

\def\muas{\mu{\rm as}}

\def\rel{{\rm rel}}

\def\rot{{\rm rot}}

\def\hel{{\rm hel}}
\def\geo{{\rm geo}}
\def\e{{\rm E}}
\def\bpi{{\bdv\pi}}
\def\bmu{{\bdv\mu}}
\def\bOmega{{\bdv\Omega}}

\def\bv{{\bf v}}

\begin{document}
\title{{\it Spitzer} Opens New Path to Break Classic Degeneracy for Jupiter-Mass Microlensing Planet OGLE-2017-BLG-1140Lb}

 \author{\textsc{
S.~Calchi~Novati$^{1}$,
J.~Skowron$^{2}$, 
Y.~K. Jung$^{3}$\\
\and 
C.~Beichman$^{4}$,
G.~Bryden$^{5}$,
S.~Carey$^{6}$,
B.~S.~Gaudi$^{7}$,
C.~B.~Henderson$^{8}$,
Y.~Shvartzvald$^{5,^{\dag}}$, 
J.~C.~Yee$^{9}$, 
W.~Zhu$^{10}$\\ 
({\it Spitzer} Team)\\
A.~Udalski$^{2}$,
M.~K.~Szyma\'{n}ski$^{2}$, 
P.~Mr\'{o}z$^{2}$, 
R.~Poleski$^{2,7}$, 
I.~Soszy\'{n}ski$^{2}$, 
S.~Koz{\l}owski$^{2}$,
P.~Pietrukowicz$^{2}$, 
K.~Ulaczyk$^{2}$,
M.~Pawlak$^{2}$,
K.~Rybicki$^{2}$,
P.~Iwanek$^{2}$\\
(OGLE Collaboration)\\
M.~D.~Albrow$^{11}$,
S.-J.~Chung$^{3,12}$, 
A.~Gould$^{3,7,13}$, 
C.~Han$^{14}$, 
K.-H.~Hwang$^{3}$, 
Y.-H.~Ryu$^{3}$, 
I.-G.~Shin$^{9}$, 
W.~Zang$^{15}$, 
S.-M.~Cha$^{3,16}$, 
D.-J.~Kim$^{3}$, 
H.-W.~Kim$^{3}$, 
S.-L.~Kim$^{3,12}$, 
C.-U.~Lee$^{3,12}$,
D.-J.~Lee$^{3}$, 
Y.~Lee$^{3,16}$, 
B.-G.~Park$^{3,12}$,
R.~W.~Pogge$^{7}$ \\
(KMTNet Collaboration)\\
} 
}

\affil{$^{1}$IPAC, Mail Code 100-22, Caltech, 1200 E. California Blvd., Pasadena, CA 91125, USA}

\affil{$^{2}$Warsaw University Observatory, Al. Ujazdowskie 4,
00-478 Warszawa, Poland}

\affil{$^{3}$Korea Astronomy and Space Science Institute, Daejon
34055, Republic of Korea}

\affil{$^{4}$NASA Exoplanet Science Institute, California Institute of Technology, Pasadena, CA 91125, USA}

\affil{$^{5}$Jet Propulsion Laboratory, California Institute of
Technology, 4800 Oak Grove Drive, Pasadena, CA 91109, USA}

\affil{$^{6}$Spitzer, Science Center, MS 220-6, California Institute of Technology,Pasadena, CA, USA}

\affil{$^{7}$Department of Astronomy, Ohio State University, 140 W.
18th Ave., Columbus, OH 43210, USA}

\affil{$^{8}$IPAC/NExScI, Mail Code 100-22, Caltech, 1200 E. California Blvd., Pasadena, CA 91125, USA}

\affil{$^{9}$Harvard-Smithsonian Center for Astrophysics, 60 Garden St., Cambridge, MA 02138, USA}

\affil{$^{10}$Canadian Institute for Theoretical Astrophysics, 
University of Toronto, 60 St George Street, Toronto, ON M5S 3H8, Canada}

\affil{$^{11}$University of Canterbury, Department of Physics and
Astronomy, Private Bag 4800, Christchurch 8020, New Zealand}

\affil{$^{12}$Korea University of Science and Technology, Daejeon 34113, Republic of Korea}

\affil{$^{13}$Max-Planck-Institute for Astronomy, K\"{o}nigstuhl 17,
69117 Heidelberg, Germany}

\affil{$^{14}$Department of Physics, Chungbuk National University,
Cheongju 28644, Republic of Korea}

\affil{$^{15}$Physics Department and Tsinghua Centre for
Astrophysics, Tsinghua University, Beijing 100084, China}

\affil{$^{16}$School of Space Research, Kyung Hee University,
Yongin, Kyeonggi 17104, Republic of Korea}

\affil{$^{\dag}$NASA Postdoctoral Program Fellow}

\begin{abstract}
We analyze the combined {\it Spitzer} and ground-based data for 
OGLE-2017-BLG-1140 and show that the event was generated by a Jupiter-class
$(m_p\simeq 1.6\,M_{\rm jup})$ planet
orbiting a mid-late M dwarf $(M\simeq 0.2\,M_\odot)$ that lies 
$D_{LS}\simeq 1.0\,\kpc$ in the foreground of the microlensed,
Galactic-bar, source star.  The planet-host projected separation is
$a_\perp \simeq 1.0\,\au$, i.e., well-beyond the snow line.  By measuring
the source proper motion $\bmu_s$ from ongoing, long-term OGLE imaging,
and combining this with the lens-source relative proper motion 
$\bmu_\rel$ derived from the microlensing solution, we show that
the lens proper motion $\bmu_l=\bmu_\rel + \bmu_s$ is consistent
with the lens lying in the Galactic disk,
although a bulge lens is not ruled out. We show that
while the {\it Spitzer} and ground-based data are comparably well fitted
by planetary (i.e., binary-lens, 2L1S) models and by binary-source (1L2S)
models, the combination of {\it Spitzer} and ground-based data decisively
favor the planetary model.  This is a new channel to resolve the
2L1S/1L2S degeneracy, which can be difficult to break in some cases.
\end{abstract}

\keywords{gravitational lensing: micro}

\section{{Introduction}
\label{sec:intro}}

The degeneracy between binary-lens/single-source (2L1S) and 
single-lens/binary-source (1L2S) microlensing events, first noted
by \citet{gaudi98}, has continually grown in importance and
complexity over the first 15 years of microlensing planet detections,
particularly as these have reached toward lower planet-host mass-ratio
planets.  As originally formulated by \citet{gaudi98}, a 1L2S event
can mimic a 2L1S event if the second source is much fainter than the
first and if the lens happens to pass much closer to it. In this case,
the second source gives rise to a smooth, short-lived, low-amplitude bump,
as it very briefly becomes highly magnified.  Any putative planetary
signal that is consistent with such a smooth short-lived bump must therefore
be vetted against the 1L2S explanation.  This already became an issue
for the third microlensing planet, OGLE-2005-BLG-390Lb \citep{ob05390},
for which the smooth bump was actually generated by a ``Cannae'' type
``Hollywood'' event \citep{ob170173}, in which a very large source
completely envelops the planetary caustic.  For the actual case of
OGLE-2005-BLG-390Lb, the 1L2S solution was ruled out ($\Delta\chi^2>50$).
However, in the course of their systematic study of all archival
low-mass-ratio $(q<10^{-4})$ microlensing planets, \citet{ob171434}
showed that had the mass ratio been smaller, $\log (q^\prime/q)<-0.2$,
then the 2L1S and 1L2S models could not have been reliably distinguished.

Over the years, it has become clear that a variety of other microlensing-planet
geometries can induce smooth bumps that can potentially be confused with
1L2S geometries.  \citet{ob161195a} and \citet{ob161195b} analyzed a
smooth bump in OGLE-2016-BLG-1195 and both showed that it was due to 
the source passing over a smooth ``ridge'' in the magnification pattern
between the central and planetary caustic (``wide planet'' solution)
or over a smooth ridge extending from the central caustic
(``close planet'' solution).  Again, however, \citet{ob171434}
showed that the 2L1S and 1L2S solutions could not have been distinguished
if the planet mass ratio had been lower by $\log (q^\prime/q)<-0.3$.

Both of these forms of the degeneracy are likely to become more important in the
future.  \citet{zhu14} showed that in the era of pure-survey microlensing
planet detections, half of all ``detectable'' planets (based on $\chi^2$
criterion) are likely to be non-caustic crossing events (i.e., broadly
similar to OGLE-2016-BLG-1195), which will generically induce smooth
bumps, as opposed to the sudden jumps that usually characterize
caustic crossings, which are present in a substantial majority of
published planetary microlensing events.  Moreover, it is not necessary
to fully envelop the caustic to produce a smooth bump in a caustic-crossing
event: \citet{ob170173} showed that ``von Schlieffen'' type Hollywood
events, in which the source only partially envelops the caustic, can
produce very similar light curves to ``Cannae'' 
events\footnote{As discussed at somewhat greater length 
by \citet{ob170373}, the
term ``Hollywood'' was coined by \citet{gould97} to emphasize the
virtues of ``following the big stars'' because they have a large
cross section for completely enveloping the planetary caustic.
Later, \citet{ob170173} distinguished between full (``Cannae'') and
partial (``von Schlieffen'') envelopment, in analogy to the military
strategies of Hannibal at Cannae and the ``von Schiefflen plan''
in World War I.}.

Furthermore, new forms of this degeneracy are being discovered.
\citet{ob160733} showed that a 1L2S event with a source-flux ratio
$q_f \simeq 2$ could be broadly mimicked by a planetary microlensing
geometry.  In this case, the 2L1S geometry was ruled out by
$\Delta\chi^2>500$, so strictly speaking the solutions were not
``degenerate''.  Nevertheless, the fact that much more complex
binary-lens structures than ``short-lived bumps'' can be mimicked
by binary-source geometries should serve as a broad caution when
analyzing events.

Finally, \citet{ob151459} found yet another path to this degeneracy
in their analysis of OGLE-2015-BLG-1459.  The first point to note
about this event is that it had a three-fold degeneracy 3L1S versus
2L2S versus 1L3S.  In the triple lens model, the third body was
a ``moon'' that was detected in only one magnified point (albeit,
a 0.4 mag deviation detected with very high confidence).  Such
single-point (or even few-point) deviations due to a planet can
easily be confused with a ``smooth bump'', even if the underlying
light curve would reveal a pronounced caustic structure, just because
of poor sampling.

The first line of defense against the 2L1S/1L2S degeneracy is
simply $\Delta\chi^2$ between the two models.  A few cases were
mentioned above, but there are many others as well (e.g., \citealt{ob170482}).
However, as discussed above, in the few published cases of low-$q$ events
that were investigated by \citet{ob171434}, the threshold for resolving
this degeneracy did not lie far below the actual value of $q$.

A second line of defense is to measure the color difference of the
(putative) two sources.  Because light travels on geodesics,
microlensing is intrinsically achromatic.  The only exception\footnote{In
fact, interference effects in microlensing (so-called ``femtolensing'',
\citealt{gould92a}) can also generate chromatic effects.  However,
this is not a practical issue for Galactic microlensing studies.}
 would
be if two stars (or two parts of a single star) were of different colors
and were magnified by different amounts.  The latter effect can occur
if a single star is transited by a point lens or by a caustic from a
binary lens.  However, this is rather weak.  Substantial chromaticity
requires two sources of substantially different color and magnified
by different amounts.  The short-term ``smooth bumps'' that are the
main source of ambiguity are well suited to this test.  Recall that
the 1L2S model generally requires that one of the sources is much
fainter than the other and also much more highly magnified.  Generally,
fainter sources are redder (particularly if the brighter source is
on the main sequence), so the light during the bump should be redder
than on the rest of the light curve.  For example, \citet{ob151459}
confirmed the 1L3S interpretation using this effect for OGLE-2015-BLG-1459.  
However, if the primary source is a giant, then the secondary can have
a similar color even if it is several orders of magnitude fainter.
Moreover, as mentioned above, there are cases for which the source-flux
ratio is actually close to unity \citep{ob160733}.  But the main
impediment to this method is simply that alternate (usually $V$) band
data are not typically taken at high-enough cadence to accurately
measure the color of a short-lived smooth bump.

Here, we use {\it Spitzer} observations of the
planetary microlensing event OGLE-2017-BLG-1140 to
demonstrate the power of a new method to resolve 
the 2L1S/1L2S degeneracy that is based on 
space-based microlensing parallax.

\section{{Observations}
\label{sec:obs}}

OGLE-2017-BLG-1140 is at (RA,Dec) = (17:47:31.93,$-$24:31:21.6)
corresponding to $(l,b)=(4.0,1.9)$.  
It was discovered and announced as a probable microlensing event
by the OGLE Early Warning
System \citep{ews1,ews2} at UT 11:57 on 19 June 2017.
The event lies
in OGLE field BLG633 \citep{ogle-iv}, for which OGLE observations
were at a characteristic cadence of $\Gamma = 1\,{\rm day^{-1}}$ 
using their 1.3m telescope at Las Campanas, Chile.

The Korea Microlensing Telescope Network (KMTNet, \citealt{kmtnet})
observed this field from its three 1.6m
telescopes at CTIO (Chile, KMTC), SAAO (South Africa, KMTS) and SSO 
(Australia, KMTA),
in its BLG19 field, implying that it was observed at a
cadence of $\Gamma = 1\,{\rm hr^{-1}}$ during the {\it Spitzer} season.
The event was identified by KMTNet as SSO19M0601.004271.

The great majority of ground-based observations were carried out in 
the $I$ band with occasional $V$-band observations made
solely to determine source colors.
All reductions for the light curve
analysis were conducted using variants of difference image analysis (DIA,
\citealt{tomaney96,alard98}), specifically \citet{wozniak2000} and \citet{albrow09}.

The event was also observed by {\it Spitzer}.  As discussed in detail
by \citet{yee15}, {\it Spitzer} selections can be ``objective'', ``subjective'',
or ``secret'', which impacts how detected planets (and planet sensitivity)
enter the {\it Spitzer} program to measure the Galactic distribution of 
planets \citep{prop2013,prop2014,prop2015a,prop2015b,prop2016}.
  Events that meet certain pre-specified objective
criteria {\it must} be observed, and consequently all planets detected during
the event can enter the program sample.  Events can be selected 
``subjectively'' by the team for any reason and at any time.  However,
only planets (and simulated planets needed to calculate planet sensitivity)
that do not generate significant signal in the data available at the
time of the public announcement can enter the sample.  The observational
cadence and the conditions for stopping the observations must be specified
at the time of the announcement.  Events can also be chosen ``secretly'',
i.e., without announcement, and then later changed to ``subjective''
(if such a decision is subsequently made).  In this case, the constraints
on what is a ``detectable'' planet apply according to the date of the
``subjective'' announcement.  Moreover, {\it Spitzer} observations taken
before this date cannot be included in the determination of whether
the microlens parallax is well-enough measured to enter the sample 
\citep{zhu17}.

OGLE-2017-BLG-1140 was chosen ``secretly'', at UT 13:08 on 19 June, i.e.,
slightly more than one hour after it was announced by OGLE and about 8 minutes
before the first {\it Spitzer} ``upload'' (i.e., when target coordinates
are sent to {\it Spitzer} Operations). 
This selection was made by the upload subteam because the event appeared
to be consistent with reaching relatively high magnification based on
the data available at that time.  The upload subteam does not generally
have the authority to choose events subjectively, without giving the
whole team an opportunity for a joint decision.
The target entered the {\it Spitzer}
Sun-angle window roughly 1.65 days after the first {\it Spitzer} observation,
i.e., at UT 07:08 on 24 June.  The event was announced ``subjectively'' at
UT 16:23 on 25 June, i.e., about 33 hours later\footnote{This decision was made
because it was realized (based on ``quick look'' KMTNet data) that the event 
would become ``objective'' 21 hours later at the next {\it Spitzer} upload.
Note that events can {\it only} become ``objective'' at the times of uploads.  
Note also that, according to the \citet{yee15} protocols as they operated
at the time of this decision, if the event had simply been ``allowed'' 
to become ``objective'' 
(i.e., without ``subjective'' announcement), then the 
{\it Spitzer}
data taken prior to the first spacecraft commands (UT 23:52, 29 June)
that were uploaded on that
date (26 June)
could not enter the \citet{zhu17} test to determine whether the parallax
had been measured well enough to enter the sample.  In fact, this is
a shortcoming of these protocols, which we now modify for future events
as follows:
if an event goes from ``secret'' to ``objective'' 
(and unless otherwise publicly specified by the team), then it automatically
becomes ``subjective'' at the upload time as well, with the cadence and conditions being
identical to those of ``objective'' events.  In this case, the usual
\citet{yee15} algorithm for resolving conflicts between ``subjective''
and ``objective'' designations is applied.  In particular, if the {\it Spitzer}
data from after the upload triggered by the ``objective'' designation
are adequate for measuring the parallax according to the \citet{zhu17}
criteria, then all planets discovered in the event can enter the sample.
However, if meeting these criteria requires earlier {\it Spitzer} data
(but still taken after the event became ``objective''), then only planets
that do not generate significant signals in data available before this
date can be included.  It may appear to be simple enough to make the
appropriate announcement on or before the date that the event becomes
``objective'' (as was done in the present case).  However, in practice,
``secret'' events receive less scrutiny during the hectic process of
evaluating hundreds of events in preparation for upload because they
do not require observing decisions.
See \citet{ob161190} for a relevant example.}.
Therefore, all {\it Spitzer}
observations in this interval must be excluded from the determination
of whether the {\it Spitzer} parallax is well measured, and of course
if a planetary anomaly proves significantly detectable $(\Delta\chi^2=10)$
from data available prior to this announcement 
(HJD$^\prime\equiv {\rm HJD}-2450000=7930.19$), then the planet must be
excluded from the Galactic distribution sample.

In fact, these restrictions have almost no practical effect.  There were
only four {\it Spitzer}
observations taken in this interval, and they do not contribute
significantly to the parallax measurement.  The last 
ground-based data point available
at the time of the announcement was at ${\rm HJD}^\prime=7928.61$, 
at which point the light curve was perfectly consistent with a point lens.
The event was first suspected to be anomalous on 5 July, but in retrospect
this appears to be based on some points near peak that were impacted
by close passage of the Moon.  The anomaly was first recognized as due
to a weak 2L1S perturbation or a 1L2S geometry on 13 July based
on ground-based data.  However, at that point, and also at a subsequent
update when the event reached baseline, the 2L1S/1L2S degeneracy
appeared insurmountable.  The decision to pursue the analysis was made
after inspecting the anomaly in the {\it Spitzer} data.

The {\it Spitzer} data were reduced using specially designed
software \citep{170event}.

We follow the standard procedure (e.g. \citealt{yee12}) 
of rescaling error bars
so that the $\chi^2/\mathrm{dof}$ for each data set
is of order unity for the best model. For OGLE and KMTNet
the rescaling factors are in the range 1.3-1.5, for \emph{Spitzer}
we evaluate a factor 2.6.

\section{{Light Curve Analysis}
\label{sec:anal}}

\begin{figure}
\epsscale{0.85}
\plotone{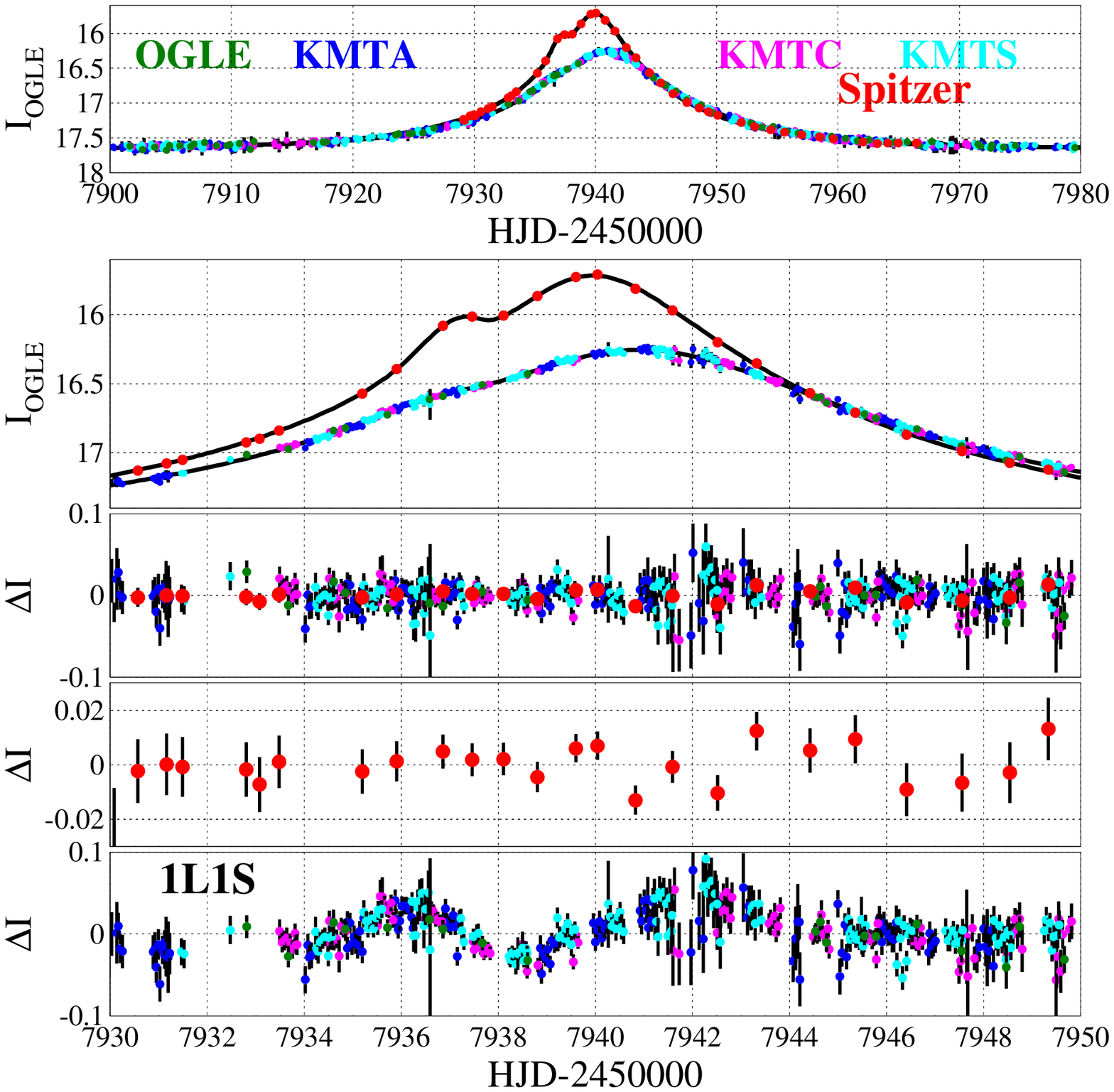}
\caption{Light curve and binary-lens/single-source (2L1S) model
and residuals
of OGLE-2017-BLG-1140 (first four panels from the top).  
The overall difference between the
{\it Spitzer} (which is transformed for display to the $I$-band
magnitude system)
and ground-based (OGLE, KMTA, KMTC, and KMTS) data
yields the microlens parallax vector $\bpi_\e$.  More subtle differences,
such as the strength of the pre-peak ``smooth bump'' anomaly in both
data sets, allow one to decisively rule out the competing class of
single-lens/binary-source (1L2S) models.  Note that the {\it Spitzer}
residuals are shown again, separately, in the bottom panel because their
error bars are substantially smaller than the range that must be displayed
on the main residual panel. Bottom panel: residuals for 1L1S
model for ground-based data only.
}
\label{fig:lc}
\end{figure}

At glance, OGLE-2017-BLG-1140 deviates
from the smooth, symmetric (single) point source-point-lens (1L1S)
\cite{pac86} shape. This is obvious from inspection
of the \emph{Spitzer} light curve, somewhat less so
for ground-based data (see Figure~\ref{fig:lc}).
Still, for ground-based data only, a 1L1S model has
a $\Delta\chi^2=915$ from the best planetary model 
discussed below, and the systematic deviations
from the data are clearly visibile in the bottom panel of Figure~\ref{fig:lc}.
Based on the general appearance of its light curve, OGLE-2017-BLG-1140
could in principle be either 2L1S or 1L2S.  However, because the correct
model is actually 2L1S, we focus on that here and defer discussion of
1L2S models to Section~\ref{sec:method}.

We will eventually show that 2L1S solutions could be derived
from either the ground-based or {\it Spitzer} data.
However, we begin by reporting our actual path toward deriving
the solution.  As in the case of OGLE-2017-BLG-1130 \citep{ob171130},
the binarity of the lens is much more apparent by eye in the {\it Spitzer}
data, so we begin by conducting a grid search using these data
only.  The lens system is reasonably well described by six parameters 
$(t_0,u_0,t_\e,s,q,\alpha)$.  The first three \citep{pac86} parameters
are, respectively the time of closest approach to the center of mass,
the impact parameter (normalized to $\theta_\e$) and the Einstein timescale,
i.e., $t_\e=\theta_\e/\mu_\rel$, where $\bmu_\rel$ 
is the lens-source relative proper
motion.  The final three are the planet-host separation 
(in units of $\theta_\e$), the planet-host mass ratio, and the angle
between the instantaneous planet-host axis and $\bmu$.  In fact, as we
will show shortly, a seventh parameter can also be measured:
$\rho\equiv\theta_*/\theta_\e$,
where $\theta_*$ is the angular radius of the source.  However, 
in order to quantify the robustness of this measurement and also to facilitate
understanding of the information flow, we initially
set $\rho=0$.  In addition to these six geometric parameters, there
are two flux parameters $(f_{s,j},f_{b,j})$ for each observatory, $j$.
That is, we model the flux observed at each time $t_i$
by the $j$th observatory as
$F_j(t_i) = f_{s,j}A(t_i;t_0,u_0,t_\e,s,q,\alpha) + f_{b,j}$.

\subsection{{Six-Parameter Solutions ($\rho=0$)}
\label{sec:rhozero}}

The binary lens model is specified 
by the parameters of the caustic, $(s,q)$, and the 
angle of the trajectory, $\alpha$. In the caustic
region, because of the
divergences in the magnification map, the $\chi^2$
topology is however extremely complex 
and may present sharp variations
along these parameters. Therefore,
standard minimization procedures including
Markov Chain Monte Carlo (MCMC)
are not suitable tools to locate, starting from
a generic position in the parameter space, the absolute minimum.
This is not the case, on the other hand,
for the single lens parameters for which
the $\chi^2$ surface is smooth
(in particular this holds for $\rho$
and explain why we run the grid
with $\rho=0$). This is the reason
why, lacking a plausible a priori intuition
of the ``right'' binary lens model,
we start the analysis with a blind-search grid
in the binary lens parameter space.
Once the $\chi^2$ minimum is identified within the grid search,
we then run a final analysis with all the parameters left free to vary
to fully characterize the solution. Specifically, we
conduct a dense $40^3$ grid search on 
$[0\leq s< 2]\times[-4\leq\log q< 0]\times[0\leq \alpha<2\pi]$,
and for each such triple we fit for the remaining three 
parameters\footnote{For the non linear fitting we make use
of MINUIT \citep{minuit75} within the CERNLIB package
https://cernlib.web.cern.ch/cernlib/.}.
We model the light curves using the algorithm of \citet{bozza10},
which has a publicly available 
implementation\footnote{http://www.fisica.unisa.it/GravitationAstrophysics/VBBinaryLensing.htm.}.
This grid search yields four minima at
$(s,\log q) =(0.7,-1.8)$, $(0.8,-2.2)$, $(1.5,-2.2)$, and $(1.6,-1.4)$.
We then seed these solutions into
a MCMC \citep{subo09} for which all parameters are allowed
to vary.  The two ``close'' ($s<1$) seeds both converge to the same 
solution, which is given in Table~\ref{tab:fixed_rho}.  
The remaining two solutions,
which are the corresponding ``wide'' $(s>1)$ variants of the close/wide 
$(s\leftrightarrow s^{-1})$ degeneracy \citep{griest98,dominik99}, also
converge.  However, these prove not be viable, as we discuss further below.

To combine {\it Spitzer} and ground-based data, we must introduce two
additional parameters, the two components of the
vector microlens parallax \citep{gould92,gould00},
\begin{equation}
\bpi_\e\equiv {\pi_\rel\over\theta_\e}\,{\bmu_\rel\over\mu_\rel}\,,
\label{eqn:piedef}
\end{equation}
where $\pi_\rel\equiv \au(D_L^{-1}-D_S^{-1})$ is the lens-source relative
parallax.  We evaluate $\bpi_\e$ in equatorial coordinates, i.e.,
$\bpi_\e = (\pi_{\e,N},\pi_{\e,E})$.

We make an initial estimate of $\bpi_\e$ by simultaneously fitting
the ground and space data (with the anomaly excised)
to a 1L1S model.  We then seed the resulting
$\bpi_\e$ as well as the {\it Spitzer}-based fit for the other six
parameters $(t_0,u_0,t_\e,s,q,\alpha)$ into a simultaneous fit to the
ground-based and {\it Spitzer} data.  The resulting solution is
again shown in Table~\ref{tab:fixed_rho}.  This is the so-called ``$(+,+)$'' solution.
See Section~\ref{sec:rhofree}.  In order to facilitate comparison
with results in that section, we also show the corresponding ``$(-,-)$''
solution.
Finally, we remove the {\it Spitzer} data
and fit for six parameters only $(t_0,u_0,t_\e,s,q,\alpha)$ using the
ground-based data.  This solution is also shown in Table~\ref{tab:fixed_rho}.

Comparing the three solutions ({\it Spitzer}-only, ground-only, and 
joint $(+,+)$), we see that they are nearly
identical.  There are only two major differences\footnote{The much
more subtle differences in $(s,q)$ are discussed in Section~\ref{sec:rhofree}.}.
First, the joint solution
has parallax parameters, whereas the others do not.  Second, the
values of $(t_0,u_0)$ for the {\it Spitzer}-only solution differ
significantly from the other two, which agree with each other.  These
two differences both reflect the fact that $\bpi_\e$ can only be determined
by comparing the ground-based and {\it Spitzer} light curves.  This means,
first, that these parameters appear only in the joint solution, and second
that the basis of the $\bpi_\e$ measurement is the different values
of $(t_0,u_0)$ as seen from the two telescope locations\footnote{Note that,
following the usual convention \citep{gould04},
the parallax parameters $\bpi_\e$ are defined in the geocentric frame
at the peak of the event as seen from Earth.  Hence, $(t_0,u_0)$
are, almost by construction. nearly identical for the ground-only and joint 
solutions.} \citep{refsdal66,gould94}.  

We also investigate the ``wide'' solutions discussed above, but find
that they are strongly excluded.  First, we repeat the entire procedure
above, but for the ground-only data.  We find seven seed solutions, of
which three converge in the MCMC to the same solution shown in Table~\ref{tab:fixed_rho}.
Of the remaining four, two converge to solutions with $\Delta\chi^2>150$,
which we consider ruled out, and the other two converge to a ``wide'' variant
from the $(s\leftrightarrow s^{-1})$ degeneracy, 
namely 
\begin{equation}
(s,\log q,\alpha)=(1.64\pm 0.02,-2.31\pm 0.04,2.487\pm 0.006)
\qquad ({\rm wide;\ ground})\,.
\label{eqn:wide-ground}  
\end{equation}
This solution already has
$\Delta\chi^2=135$ relative to the ground-only solution in Table~\ref{tab:fixed_rho}.
However, the main thing to note is that the $(s,\log q,\alpha)$ parameters are
different from those reported from the {\it Spitzer}-only ``wide'' solution
discussed above, which stated more precisely are
\begin{equation}
(s,\log q,\alpha)=(1.57\pm 0.02,-2.03\pm 0.04,2.575\pm 0.010)
\qquad ({\rm wide;}\ Spitzer)\,.
\label{eqn:wide-spitzer}  
\end{equation}
This discrepancy is related to the fact that, at next order in $q$
(i.e., away from the $q\rightarrow 0$ limit), the 
$(s\leftrightarrow s^{-1})$ degeneracy 
is actually trajectory-specific \citep{an05}.
That is, it becomes a one-dimensional (1-D) degeneracy on a cut through
the 2-D magnification plane.  See Figure~4 from \citet{mb9947}
and Figure~8 from \citet{ms9801}.  Hence, when both ground-based and 
{\it Spitzer} data sets are fit jointly to the ``wide'' solution,
they prove incompatible, with $\Delta\chi^2=522$ (compared to 2L1S),
i.e., 358 higher than the sum of the two $\Delta\chi^2$ from the separate fits.

\subsection{{Seven-Parameter Solutions (Free $\rho$)}
\label{sec:rhofree}}

\begin{figure}
\epsscale{0.95}
\plotone{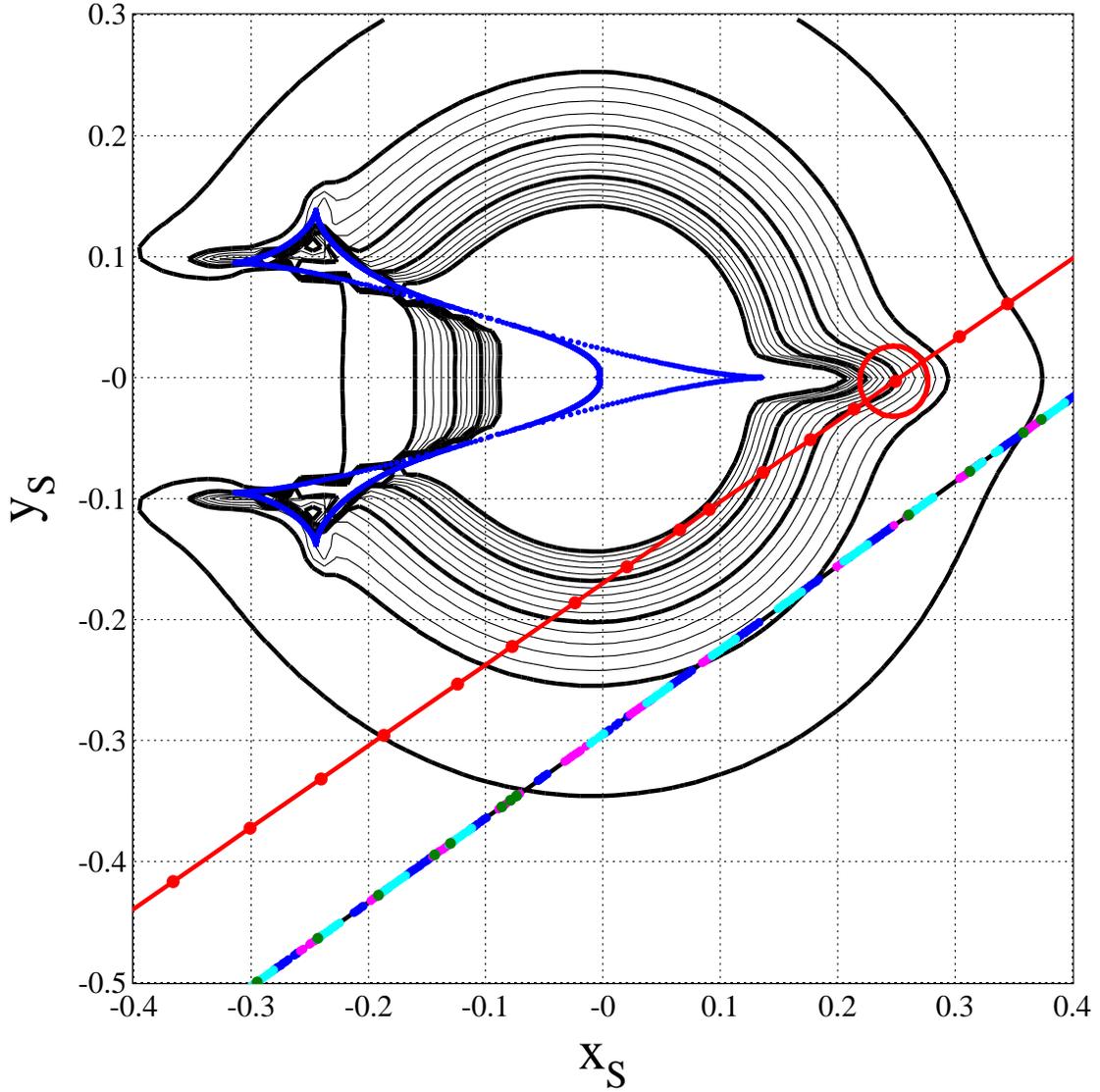}
\caption{Lens geometry for the ``$(+,+)$'' 2L1S model of 
OGLE-2017-BLG-1140.  The caustic structure is shown by a closed
concave polygon. The point-source magnification contours for 
$A_{\rm point-source}=(3,4,5,6,7)$ are shown in thick lines, with finer grading
shown in thin lines.
The two source trajectories (space and ground)
are populated by source positions (relative to the lens structure)
at the times of observations.  These are color-coded by observatory.
The source size is shown as an open red circle.  This illustrates
how the source is resolved by the ``magnification ridge'' that extends
from the cusp along the $x$-axis.
}
\label{fig:caust}
\end{figure}

Next, we allow $\rho$ to vary freely in the MCMC, seeded by the
{\it Spitzer}-only, ground-only, and joint $(+,+)$ solutions
from Table~\ref{tab:fixed_rho}\footnote{
For the finite size source calculation, limb-darkening 
may in principle be taken into account however
the solution turns out to be in a region
of the $\rho,\,s,\,q$ parameter space, 
namely with the source always passing far enough from the caustic, 
where it has no significant effect.}
The best-fit parameters 
are shown in Table~\ref{tab:free_rho}, and the geometry of the
joint solution is shown in Figure~\ref{fig:caust}.
The ``bump'' in the {\it Spitzer} light curve is caused by the
source passing over the ridge extending from a cusp of the central
caustic.  The ground-based light curve is also affected by this
cusp passage, but because the source lies further from the cusp
as seen from Earth, its effect on the light curve is not as easily
discernible by eye.  Nevertheless, as demonstrated by the
similarity of the solutions in Tables~\ref{tab:fixed_rho} and \ref{tab:free_rho}, the 
ground-based light curve is sufficiently impacted to measure the planetary 
parameters.

The geometry shown in Figure~\ref{fig:caust} is of the so-called ``$(+,+)$''
solution, i.e., with $u_0>0$ for both ground-based and {\it Spitzer}
observatories\footnote{See Figure~4 from \citet{gould04} for the definition
of the sign of $u_0$.}.  For 1L1S parallaxes, there is a generic four-fold
degeneracy corresponding to the four possible sign combinations as seen from
Earth and the satellite, i.e., $(+,+)$, $(+,-)$, $(-,+)$, and $(-,-)$.
These can also be expressed as $(+,-)\times({\rm same},{\rm opposite})$,
where the first component gives the sign of $u_0$ as seen from Earth and
the second tells whether the satellite $u_0$
has the ``same'' or ``opposite'' sign.  For well-covered binary lenses,
we expect that the ``$({\rm same},{\rm opposite})$'' degeneracy will be
broken, although if good coverage is lacking, this degeneracy may persist
\citep{ob141050}.  Figure~\ref{fig:caust} illustrates this principle
very well.  We can see that if the Earth trajectory were transposed
to the opposite side of the host (but with the same direction), it
would be impacted by several cusps and caustics, so that its magnification
profile would completely fail to match the observed light curve.
Indeed, we confirm by numerical modeling that there are no viable ``opposite''
[$(+,-)$ and $(-,+)$] solutions.  However, there is a competitive $(-,-)$
solution, the parameters of which are given in Table~\ref{tab:free_rho}.  As is often
the case \citep{ob09020}, these parameters are nearly the same as for
$(+,+)$ except for the sign reversals of $(u_0,\alpha,\pi_{\e,N})$.

Comparing Tables~\ref{tab:fixed_rho} and \ref{tab:free_rho}, we see that there is $\Delta\chi^2=22.9$
improvement for the $(+,+)$ solution when adding $\rho$ as a free
parameter (and $\Delta\chi^2=24.9$ for $(-,-)$).  
The physical origin of this measurement lies in the narrowness of the
magnification ridge that extends from the cusp seen in Figure~\ref{fig:caust},
which is of the same order as the normalized source size.  This is 
qualitatively similar to the case of OGLE-2016-BLG-1195Lb
\citep{ob161195a,ob161195b}.  Because $\rho$ is
not constrained at all in the ground-only models (see Table~\ref{tab:free_rho}),
one might suspect that the $\chi^2$ improvement comes entirely from
the {\it Spitzer} data.  In fact, this is not the case: For the
$(+,+)$ solution, only $\Delta\chi^2_{spitzer} = 11$ comes from {\it Spitzer}
with the rest coming from the ground.  Comparing Tables~\ref{tab:fixed_rho} and \ref{tab:free_rho}, we see 
that the $(s,q)$ values for {\it Spitzer}-only and ground-only agree 
significantly better in the latter than the former.  Moreover the $(s,q)$
values of the joint solution in Table~\ref{tab:free_rho} are nearly identical to those
of the ground-only solution.  This means that the ground-only model in 
Table~\ref{tab:fixed_rho} has been forced away from its ``preferred'' solution by the
necessity to accommodate adjustments in $(s,q)$ that are needed to
reconcile the $\rho=0$ model to the {\it Spitzer} data.  Once $\rho$
is set free in Table~\ref{tab:free_rho}, the {\it Spitzer}-only model comes much
closer to the $(s,q)$ preferred by the ground-only model.  In brief,
the ground-based data acts to ``enforce'' $(s,q)$, and this indirectly
places constraints on $\rho$.  This leads to a factor $\sim 2$ reduction
in the error on $\rho$ of the joint solution compared to the {\it Spitzer}-only
solution, despite the fact that the ground-based data contain no direct
information about $\rho$.

\section{{Physical Parameters}
\label{sec:phys}}

Because $\pi_\e$ and $\rho$ are both measured, it is only necessary
to determine $\theta_*$ in order to measure the physical properties
of the system.  We will then obtain $\theta_\e = \theta_*/\rho$ and 
thereby the lens mass $M$ and lens-source relative parallax $\pi_\rel$,
\begin{equation}
M= {\theta_\e\over\kappa\pi_\e},\qquad \pi_\rel = \theta_\e\pi_\e\,,
\label{eqn:mpirel}
\end{equation}
where $\kappa\equiv {4G/c^2\au}\simeq 8.144
\,\textrm{mas}\,\textrm{M}_\odot^{-1}$ (see e.g. \cite{gould00}
for an introduction to the concepts and formalism of microlensing).

\subsection{{Information From Microlens Parallax Only}
\label{sec:pieinfo}}

Nevertheless, it is instructive to ask what can be known without
the $\theta_\e$ measurement, particularly because, for the overwhelming
majority of the non-planetary ``comparison sample'' needed to determine
the Galactic distribution of planets, $\theta_\e$ is not measured
\citep{zhu17}.

We begin by calculating the heliocentric projected velocity for the
two solutions in Table~\ref{tab:free_rho} (see also Table~\ref{tab:phys_par}, below),
\begin{equation}
\tilde \bv_\hel = \tilde \bv_\geo +\bv_{\oplus,\perp}
\equiv {\au\bpi_\e\over \pi_\e^2 t_\e} + \bv_{\oplus,\perp}\,,
\label{eqn:vtilde}
\end{equation}
which, in equatorial coordinates, can be evaluated,
\begin{equation}
\tilde \bv_\hel(N,E) = \Biggl[{(+1031,+719)\atop(-1031,+728)}\Biggr]\,\kms
\quad \Biggl[{(-,-)\atop(+,+)}\Biggr]\,.
\label{eqn:vtildehel}
\end{equation}
Here, $\bv_{\oplus,\perp} = (-0.8,+28.0)\,\kms$ is the velocity of Earth at the
peak of the event, projected onto the plane of the sky. It is notable that
the direction of the $(-,-)$ solution (i.e., $35^\circ$ north through east)
is very similar to the direction of Galactic rotation.  This would make
it highly compatible with a disk lens.  That is, in general,
\begin{equation}
\tilde \bv_\hel = {\au\over \pi_\rel}\bmu_\hel\,,
\label{eqn:vtildemu}
\end{equation}
and so, ignoring the peculiar motions of the source, lens, and Sun, we expect 
that the projected velocity will lie almost exactly in the direction
of Galactic rotation.  This is because the Local Standard of Rest
(of the Sun) and the local standards of rest of other disk stars both
partake of this motion, while the Galactic bar 
(the presumed home of the source) rotates in very nearly the same direction.

In fact, although this mean motion of the bar is usually ignored
(but see \citealt{ob160693}), this is not strictly permissible in the
present case because the Galactic longitude $l=4.0$ is relatively high.
Applying the Law of Sines and the Exterior Angle Theorem, 
one finds that for solid
body rotation at $\bOmega$, the mean source proper motion is given by
\begin{equation}
\langle\bmu_s\rangle = \sin l (\cos l \cot\psi - \sin l)\bOmega
\rightarrow (\sin l \cot\psi)\bOmega\,,
\label{eqn:barpm}
\end{equation}
where $\psi$ is the bar angle and where we have eliminated second-order
terms in the final expression.  Adopting $\Omega=75\,\kms\,\kpc^{-1}$
and $\psi=40^\circ$, we obtain $\langle\mu_s\rangle= 1.3\,\masyr$ for
this field.  Therefore, for disk lenses we expect
\begin{equation}
\langle\bmu_\rel\rangle = {\bv_\rot\over D_S} - (\sin l \cot)\psi\bOmega
\rightarrow 4.3\,\masyr\,{\bv_\rot\over v_\rot}\,,
\label{eqn:bmuest}
\end{equation}
where $\bv_\rot$ is the velocity of Galactic rotation, 
$v_\rot = |\bv_\rot|\sim 220\,\kms$, and
$D_S\sim 8.1\,\kpc$ (see Section~\ref{sec:physeval}).
Thus, while the {\it direction} of the lens-source relative motion
of the $(-,-)$ solution (Equation~(\ref{eqn:vtildehel})) favors
disk lenses, the {\it amplitude} of the expected relative proper motion
is actually very similar for both disk and bulge lenses.

Next, we insert this estimate of $\mu_\rel$ for disk lenses and the
value of $\tilde v$ from Equation~(\ref{eqn:vtildehel}) to obtain
\begin{equation}
\langle\pi_\rel\rangle = 
{v_\rot/D_S-(\sin l\cos \psi)\Omega\over\tilde v_\hel}\au 
\rightarrow 0.016\,\mas \,.
\label{eqn:pirelpre}
\end{equation}
This means that the $(-,-)$
projected velocity is very nearly what would be expected
for a disk lens with source-lens separation $D_{LS}\sim 1.0\,\kpc$.  In this
case (and taking account of Equation~(\ref{eqn:mpirel})), we should have
$\theta_\e\simeq 0.17\,\mas$ and so 
$M=\theta_\e/\kappa\pi_\e\simeq 0.22\,M_\odot$.  On the other
hand, both solutions in Equation~(\ref{eqn:vtildehel}) are quite compatible
with the lens being in the bulge, in which case $\pi_\rel$ would likely be
slightly smaller, implying (at fixed $\pi_\e$), smaller 
$M$ and $\theta_\e$ as well.

These arguments imply that, in the absence of any information about
$\theta_\e$ (the typical case for the non-planetary ``comparison sample''),
the microlens parallax measurement by itself would
not discriminate well between bulge and disk lenses.  This would
not be particularly troubling for the comparison sample because
it is used only to construct a comparison cumulative distribution
of lens distances, so the role of any particular lens in this relatively
large sample is quite minor \citep{21event,zhu17}.

However, it also shows that unless the measured lens-source relative proper motion
turns out to be unexpectedly low (which would favor a bulge lens),
this proper motion measurement is unlikely, by itself, to add to the discriminatory
power to what can be determined from the $\bpi_\e$ measurement alone.  
We return to this point in 
Section~\ref{sec:diskbulge}.

\subsection{{Color-Magnitude Diagram}
\label{sec:cmd}}

\begin{figure}
\plotone{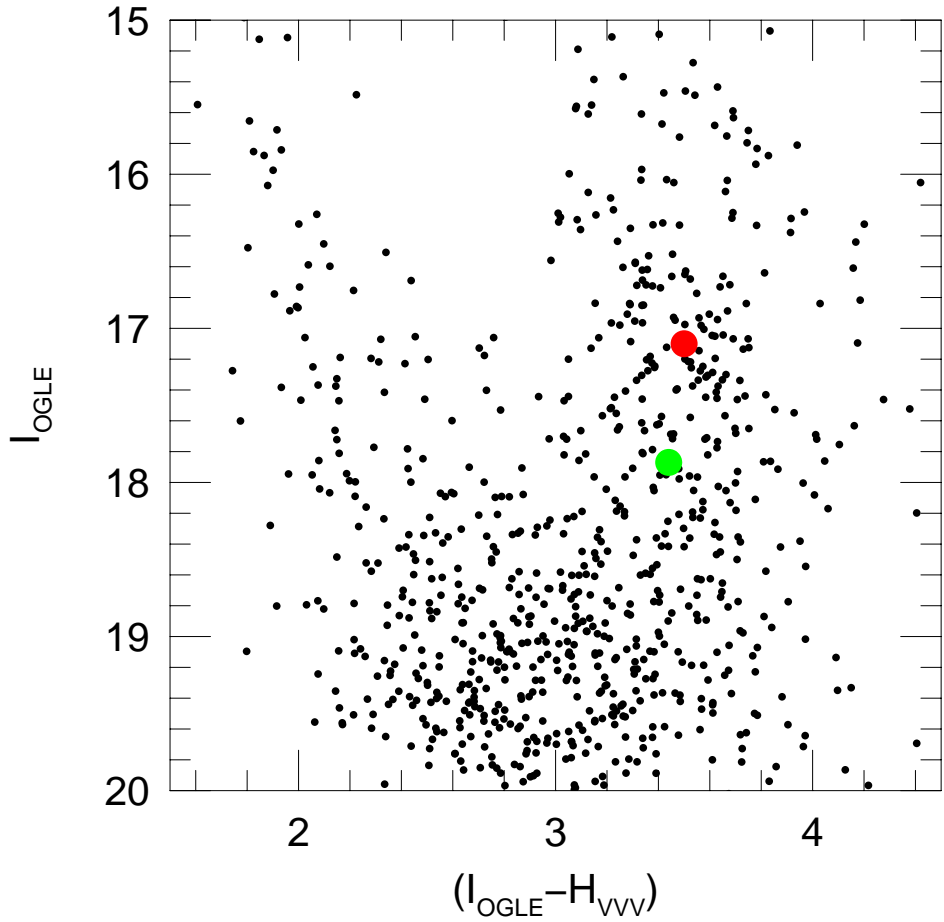}
\caption{Color-magnitude diagram (CMD) from combining OGLE-IV $I$-band
and VVV $H$-band data.  The source position (green) in these two
bands is determined from the best-fitting model to the OGLE $I$ and
SMARTS ANDICAM $H$, with the latter transformed to the VVV system
from field stars.  The clump centroid is shown in red.}
\label{fig:cmd}
\end{figure}

To measure $\theta_\e=\theta_*/\rho$, we evaluate $\theta_*$ by placing
the source on a color-magnitude diagram (CMD) \citep{ob03262}.
However, because of high extinction, $V_s$ is poorly measured,
so we cannot place the source directly on an $[I,(V-I)]$ CMD.
Instead we use SMARTS (1.3m) ANDICAM $H$-band data (together
with OGLE $I$-band data) to derive 
$(I_{\rm OGLE-IV} - H_{s,\rm ANDICAM})=-1.21\pm 0.01$ in the instrumental
system, by aligning these to the best fit model.  We then calibrate
this to the much deeper VVV catalog and find 
$H_{\rm ANDICAM}-H_{\rm VVV}=4.65\pm 0.01$ from field stars.  This yields
$(I_{\rm OGLE-IV} - H_{s,\rm VVV})=3.44\pm 0.02$.  From the fit to the light curve 
(Table~\ref{tab:free_rho}), $I_{s,\rm OGLE-IV} = 17.86\pm 0.02$.  
We compare these values to those
of the clump on the OGLE-IV/VVV CMD (Figure~\ref{fig:cmd}),
$[(I-H),I]_{\rm cl}=(3.50,17.10)\pm (0.05,0.08)$, and derive an
offset $\Delta [(I-H),I]=(-0.06,+0.76)\pm (0.05,0.08)$.
Using the color-color relations of \citet{bb88}, we translate this
to an offset $\Delta [(V-I),I]=(-0.04,+0.76)\pm (0.05,0.08)$ on the $V/I$ CMD. 
We adopt $[(V-I),I]_{0,\rm cl} = (1.06,14.33)$ from \citet{bensby13} and
\citet{nataf13}, 
to finally derive $[(V-I),I]_{0,s} = (1.02,15.09)\pm (0.05,0.08)$.  We then
convert from $V/I$ to $V/K$ using the color-color relations of \citet{bb88},
with $(V-K)=2.36\pm 0.12$.
Finally, we use the \cite{kervella04} surface brightness relation
to evaluate the angular diameter, $\theta_\mathrm{LD}$
\begin{equation}
\log(\theta_\mathrm{LD})=0.2672\,(V-K)+0.5354 - 0.2\,V\,.
\label{eqn:kervella04}
\end{equation}
The resulting source angular radius is
\begin{equation}
\theta_* = 4.39\pm 0.38\,\muas\,.
\label{eqn:thetaeeval}
\end{equation}

We recall that, within this evaluation, based
on the determination of the offset of the source 
within the CMD to the clump, the CMD itself does not
need to be calibrated, specifically zero point offsets
cancel out in the calculation. We also recall 
that the OGLE-IV $I$ bandpass is extremely close
to Cousins \citep{ogle-iv}. In particular the color term is well below
the uncertainty of measurement 
of the clump centroid\footnote{Although this is not used in the evaluation
we note that $I_\mathrm{OGLE-IV}=I_C+0.094$.}.

The final error budget for $\theta_*$, relative error 8.7\%, 
is dominated by the uncertainty
in centroiding the clump  and the conversion
$(V-I,I)$ to $(V-K,K)$. Specifically, the error in the conversion of $(I-H)$ 
to $(V-I)$ is about 0.006~mag (this is because the offset
from the clump is only 0.06~mag), well below 
the error in centroiding the clump, and so can be ignored.
The relative error would drop to 3.2\%
if we neglected the error in  centroiding the clump
and to about 5.4\% if we neglected the propagation 
error from $(V-I)$ to $(V-K)$.
Finally, the error in the surface brightness
relation is also negligible, with the relative error at the 1\% level.

\subsection{{Evaluation of Physical Parameters}
\label{sec:physeval}}

Inserting the measurements of $\rho$ and $t_\e$ from Table~\ref{tab:free_rho}, 
the value in Equation~(\ref{eqn:thetaeeval}) yields
\begin{equation}
\theta_\e = {\theta_*\over\rho} = 0.16\pm 0.02\,\mas\,,
\qquad
\mu_\rel = {\theta_\e\over t_\e} = 4.1\pm 0.6\,\masyr\,.
\label{eqn:thetaemueval}
\end{equation}
These values are very similar to those ``predicted'' 
in Section~\ref{sec:pieinfo} for a disk lens prior to incorporating
information about $\theta_\e$.  As discussed there, this immediately
implies that, although the lens distance is well measured,
we cannot, on the basis of the microlensing solution alone, 
strongly discriminate between the lens lying in the disk or the bulge.
We return to this problem in Section~\ref{sec:diskbulge}.

The simultaneous measurement of both the microlens parallax,
$\pi_\e$, and the Einstein angular radius, $\theta_\e$,
together with that of the microlensing parameter $q=M_{\rm planet}/M_{\rm host}$,
finally allow us to determine the physical parameters 
of the system (Equation~(\ref{eqn:mpirel})).
In Table~\ref{tab:phys_par} 
we present the solution, and in particular we find
\begin{equation}
M_{\rm host} = 0.21\pm 0.03\,M_\odot\,;
\qquad
M_{\rm planet} = 1.6^{+0.4}_{-0.3}\,M_{\rm jup }\,.
\label{eqn:masses}
\end{equation}
Note that in lieu of the lens distance, $D_L$, we rather report
\begin{equation}
D_{8.3}\equiv {\kpc\over 1/8.3 + \pi_\rel/\mas}\,.
\label{eqn:d8.3}
\end{equation}
The primary reason for this is that $D_{8.3}$ is much better constrained
than $D_L$ because the error in the distance to the source (due to the
finite depth of the bar) is of the same order as the distance from
the lens to the source, $D_{LS}\equiv D_S - D_L$.  Note that for cases
like the present one, for which $D_{LS}\ll D_L$, we have approximately
$D_{LS}\simeq 8.3\,\kpc - D_{8.3}$.  In particular, \citet{21event} introduced
$D_{8.3}$ in order to put all {\it Spitzer} lenses on a homogeneous
distance scale with minimal error.

However, we should also note that, at $l=4.0$, the source is fairly
far out on the near side of the Galactic bar and that the value
of $I_{\rm cl}=14.33$ adopted in Section~\ref{sec:cmd} corresponds
to a mean distance to the bar of $D_{\rm bar}\sim 7.8\,\kpc$.  If the lens
lay well in the foreground of the bar, then this would also be
a good mean estimate for $D_S$.  However, because the lens is either
in or near the bar, the mean estimate of the source distance
is ``pushed back'', simply because the cross section for
lensing scales $\sim \sqrt{D_{LS}}$.  In particular, if the lens were
known to be in the bar, the best estimate of the source distance would
be $D_S= D_{\rm bar} + D_{LS}/2=8.3\,\kpc$.  A similar effect (but not
as strong) applies to disk lenses near the bar, $D_S\sim 8.0\,\kpc$.  
We adopt $D_S\simeq 8.1\,\kpc$ to evaluate the planet-host projected 
separation, $a_\perp$,
\begin{equation}
a_\perp = 1.02\pm 0.15\,\au \,.
\label{eqn:aperp}
\end{equation}

\subsection{{Source Proper Motion}
\label{sec:diskbulge}}

We are fortunate that the source is a giant star that is 
relatively bright (despite significant extinction), relatively isolated,
and only slightly blended.  This means that we can measure the source
proper motion $\bmu_s$, which will enable a much more precise 
determination of the lens proper motion, $\bmu_l = \bmu_s+\bmu_\rel$,
than would otherwise be possible.  This can in principle provide a decisive
kinematic discriminant between the bulge-lens and disk-lens interpretations.
More specifically, as we will show, certain values of $\bmu_l$
would decisively rule out disk lenses, but no measured value of $\bmu_l$
would by itself decisively confirm the lens as belonging to the disk.

\begin{figure}
\epsscale{0.95}
\plotone{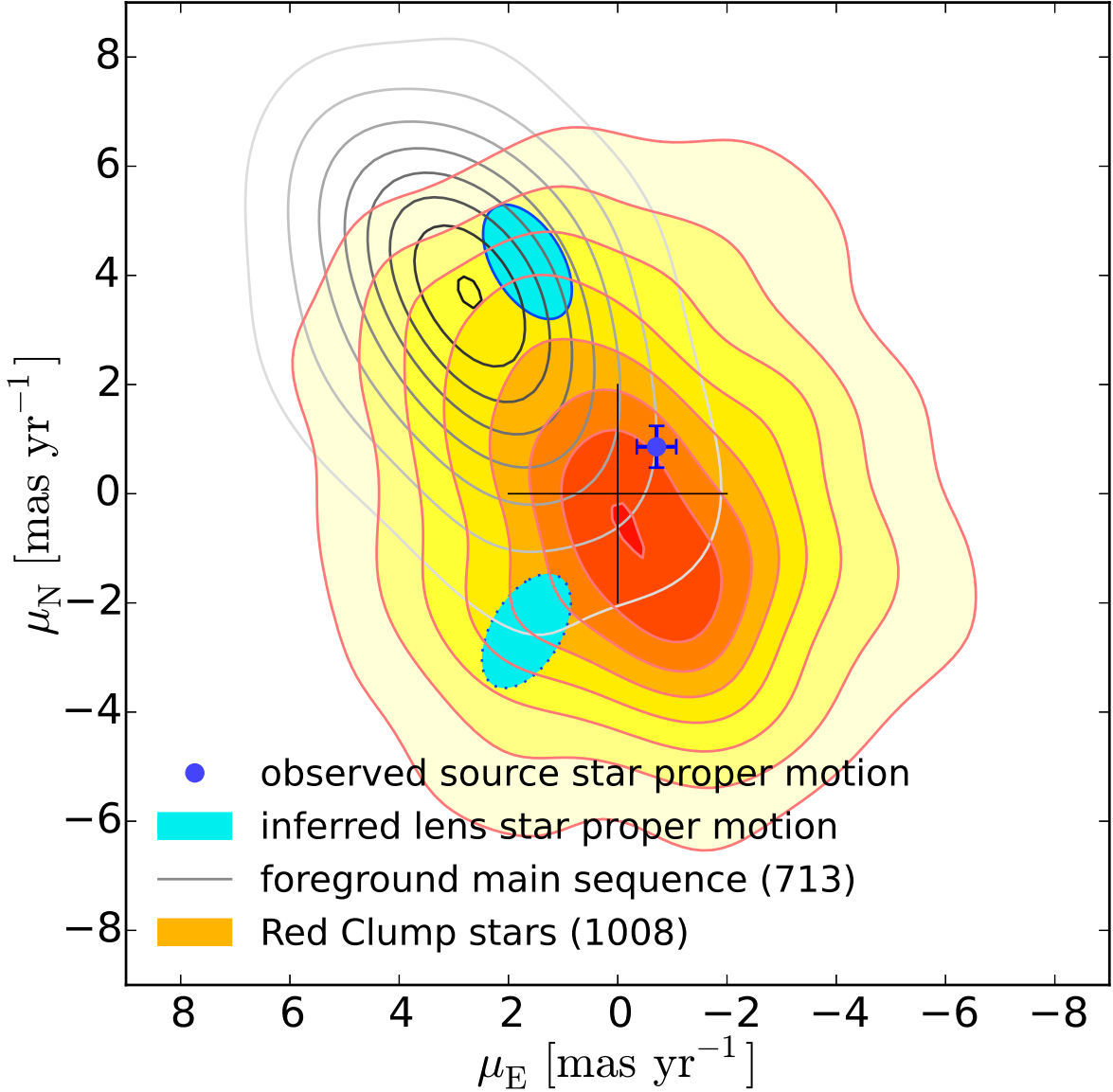}
\caption{Smoothed proper motion distributions of Galactic-bar
red clump (RC) stars and foreground disk main-sequence (MS) stars.
The source proper motion $\bmu_s$ is well-measured (blue point).  Combining
this with the two microlensing solutions in Table~\ref{tab:free_rho}
yields two possible estimates for the lens proper motion $\bmu_l$ (centers of
cyan ellipses).  The ellipses themselves take account of both the
measurement errors entering into the determination of $\bmu_l$ and
the intrinsic proper-motion dispersion of disk lenses.  See text for details.
The northern and southern ellipses correspond to the $(-,-)$ and
$(+,+)$ solutions, respectively.
}
\label{fig:smoothpm}
\end{figure}

The rationale of the present analysis is to compare
the estimated lens proper motion to that of the field
bulge and main sequence (disk) populations. 
To estimate the relative probability of a disk
versus a bulge lens, going beyond a possible ``at glance'' analysis 
from Figure~\ref{fig:smoothpm}, we should take into account both
the underlying kinematic
and density distributions for both populations. 
Below we make a detailed evaluation of
the relative probability based on the kinematic distributions.
On the other hand, the disk and bulge density profiles at the lens distance toward this
direction are too poorly understood at present to evaluate the
density term of the relative probability.
In the next few years we may expect,
following the GAIA DR2 release and therefore
the knowledge of the astrometry of the bulge as a whole,
to understand much better these density profiles.
Together with the analysis presented here, this
will then allow one to obtain a reliable estimate of
bulge-vs-disk lens.

We begin by identifying three sets of stars from a color-magnitude diagram
of stars in a $6.5^\prime$ square centered on the event: 1008 bulge
red clump (RC) stars, 2123 bulge red giant branch (RGB) stars,
 and 713 foreground main-sequence (MS) stars.  We measure
the vector proper motions of each star (relative to a frame set by
the RC stars) based on 250 (out of 708) better-seeing 
($0.9^{\prime\prime}<{\rm FWHM}<1.3^{\prime\prime}$)
OGLE-IV images from
$5275.9\leq {\rm HJD}^\prime\leq 8019.6$.
The typical proper motion errors (derived from internal scatter) are
$\sigma_\mu\sim 0.5\,\masyr$.  We exclude a handful of stars with
individual errors $\sigma>2\,\masyr$.  (The numbers given above already
take account of this exclusion.)\ \ Figure~\ref{fig:smoothpm} shows
contours of the RC and MS proper motion distributions based on smoothed
counts, and also shows the proper motion of the source star:
\begin{equation}
\bmu_s(N,E) = (0.86,-0.71)\pm(0.38,0.36)\,\masyr\,.
\label{eqn:sourcepm}
\end{equation}
Figure~\ref{fig:smoothpm} also shows the lens proper motion $\bmu_l$
together with an error ellipse (defined by
covariance matrix $c_{ij}$, which we describe further below),
\begin{equation}
\bmu_l(N,E) = (4.21,1.63)\,\masyr;
\qquad
c_{ij} = (1.10,0.42,0.42,0.67),(\masyr)^2\,,
\label{eqn:lenspm}
\end{equation}
where we have included the very small $(<0.1\,\masyr)$ correction
from geocentric to heliocentric proper motion.

From Figure~\ref{fig:smoothpm}, one sees that the lens proper motion
is offset from the peak of the MS distribution by
\begin{equation}
\Delta\bmu(N,E) = \bmu_l - \bmu_{\rm peak-MS} = (0.56,-0.89)\,\masyr\,,
\label{eqn:deltamu}
\end{equation}
where $\bmu_{\rm peak-MS}(N,E) = (3.65,2.52)\,\masyr$ is the peak of the
MS distribution.  To assess the level of consistency represented
by this offset, we consider three sources of uncertainty.  Two of these 
are error terms related to the {\it measurement} of 
$\bmu_l=\bmu_s+\hat{\bf n}\mu_\rel$,
where $\hat{\bf n}\equiv \bmu_\rel/\mu_\rel$ is the direction of $\bmu_\rel$,
i.e., the same as the direction of $\tilde\bv$.  From 
Equation~(\ref{eqn:sourcepm}), the first-term covariance matrix
is almost isotropic. On the other hand, because $\hat{\bf n}$ is measured
extremely well (see Table~\ref{tab:free_rho}), the covariance matrix
associated with $\bmu_\rel$ is nearly degenerate.  Adding these two,
we find $c^{\rm meas}_{ij} = (0.35,0.14,0.14,0.23)\,(\masyr)^2$.

The third source of uncertainty relates to the prediction of the
lens proper motion under the assumption that the lens is in the disk.
We assume that the velocity dispersion of the lenses is 
$(33,18)\,\kms$ in the rotational and vertical directions, i.e.,
similar to local disk stars.  We then rotate to equatorial coordinates
to obtain a covariance matrix 
$c^{\rm pred}_{ij} = (0.75,0.28,0.28,0.43)\,(\masyr)^2$.  We can then
evaluate the $\chi^2$ of the measured offset $\Delta\bmu$ given
these uncertainties
\begin{equation}
\chi_{\rm offset}^2= \sum_{ij} b_{ij}(\Delta\mu)_i(\Delta\mu)_j = 2.72 ;
\qquad 
b\equiv c^{-1} ;
\qquad
c_{ij} = c_{ij}^{\rm meas} + c_{ij}^{\rm pred}\,;
\label{eqn:chi2}
\end{equation}
For a 2-D Gaussian, this has probability 
$P(\chi^2_{\rm offset})=\exp(-\chi^2/2)=0.26$
which is quite reasonable.   From Figure~\ref{fig:smoothpm}, it is
clear that the great majority of stars drawn randomly from the bulge
population would have dramatically lower $P$ values.

We note that, properly speaking, the $c^{\rm meas}_{ij}$ ellipse
should be drawn around $\bmu_l$ while the $c^{\rm pred}_{ij}$ ellipse
should be drawn around $\bmu_{\rm peak-MS}$.  However, we have combined
the two covariance matrices (Equation~(\ref{eqn:chi2})) for three
reasons.  First, from a mathematical standpoint, Equation~(\ref{eqn:chi2})
remains valid regardless of whether the contributing covariance matrices
are summed before or after display.  Second, with this display, the
level of discrepancy is directly manifest in the diagram.  Third,
this mode of display will facilitate numerical evaluations below.

We also show a second error 
ellipse\footnote{$\bmu_l(N,E)=(-2.51,1.67)\,\masyr$;
$c_{ij} =(1.10,-0.42,-0.42,0.66)\,(\masyr)^2$}
in the lower part of 
Figure~\ref{fig:smoothpm}.  Any lens star that actually lay in this
ellipse would (due to the $(+,+)\leftrightarrow(-,-)$ degeneracy:
see Table~\ref{tab:free_rho}) produce the same solutions
as a corresponding star in the upper ellipse.  Hence, the two
groups of potential lenses can only be distinguished at the
$1\,\sigma$ level, and so must both be considered.

Assuming that the proper motions of lenses and sources are independent
of their distances within the narrow limits permitted by the
microlensing solution (a point to which we return below), the
relative probability of a disk versus bulge lens can be factored,
\begin{equation}
{P_{\rm disk}\over P_{\rm bulge}} = 
\biggl({P^{\rm kin}_{\rm disk}\over P^{\rm kin}_{\rm bulge}}\biggr)
\biggl({P^{dens}_{\rm disk}\over P^{\rm dens}_{\rm bulge}}\biggr);
\qquad
{P^{\rm kin}_{\rm disk}\over P^{\rm kin}_{\rm bulge}}=
{f^{(-,-)}_{\rm disk} + Q f^{(+,+)}_{\rm disk}\over f^{(-,-)}_{\rm bulge} + Q f^{(+,+)}_{\rm bulge}}\,,
\label{eqn:pdiskbulge}
\end{equation}
where $f_{\rm disk}(\bmu_l)$  and $f_{\rm bulge}(\bmu_l)$ are the normalized
proper motion distributions of the disk and bulge populations respectively
(convolved with measurement errors, as above),
$f^{(-,-)}$ and $f^{(+,+)}$ are values of these distribution at the measured
values of the two solutions, and 
$Q\equiv \exp(-(\chi^2_{{\rm mod}(+,+)}-\chi^2_{{\rm mod}(-,-)})/2)=0.61$
is the relative likelihood of the microlensing models based on
the $\chi^2$ values in Table~\ref{tab:free_rho}.  We focus here on the
first (kinematic) term, which is written more explicitly in the second
expression of Equation~(\ref{eqn:pdiskbulge}).

\begin{figure}
\epsscale{0.95}
\plotone{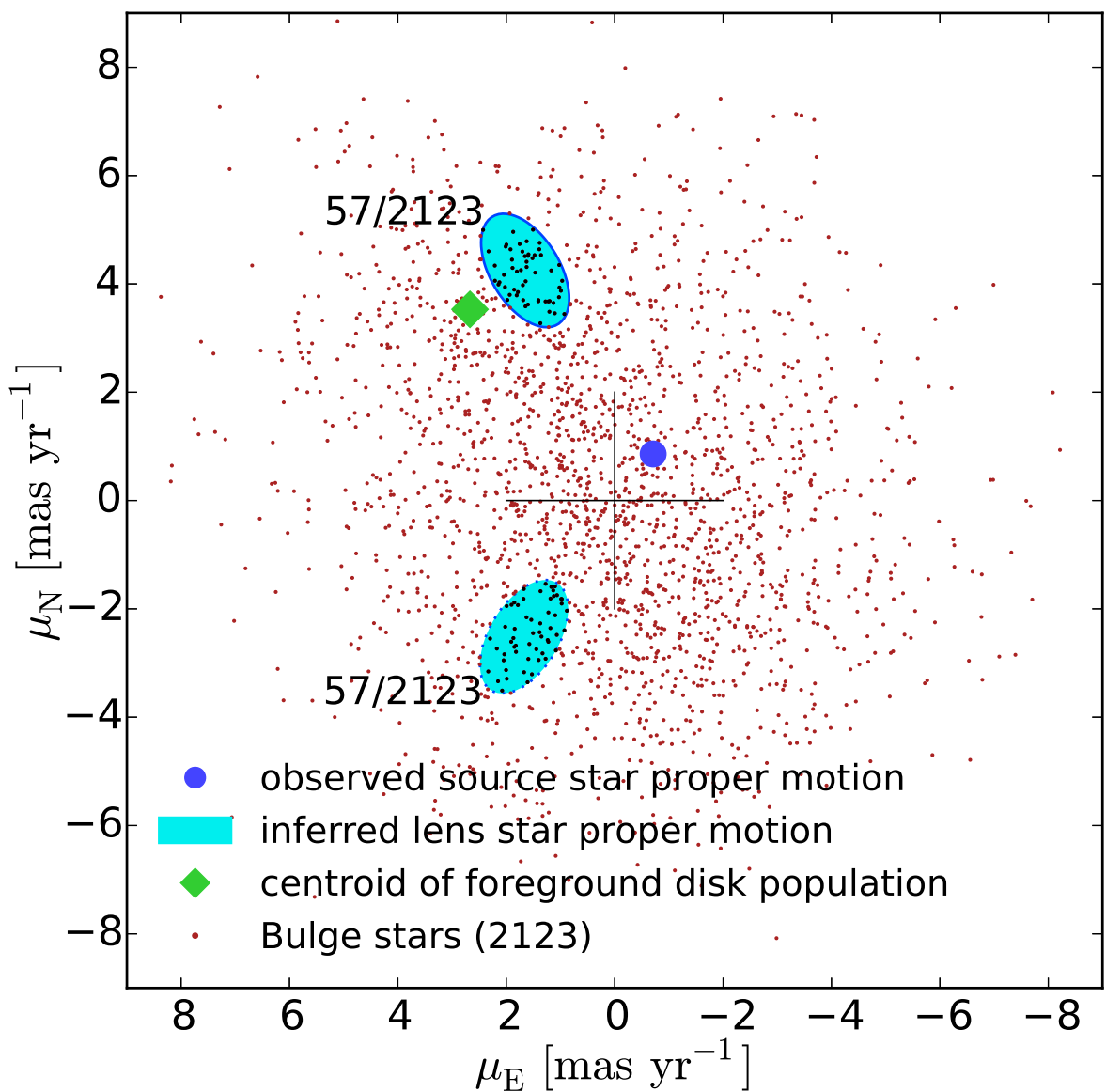}
\caption{Observed proper motions of bulge red giant branch (RGB) stars
in a $6.5^\prime$ square around OGLE-2017-BLG-1140.  The cyan ellipses
are the same as in Figure~\ref{fig:smoothpm}.  The fractions
of RGB stars that lie in each ellipse (57/2123 in both cases) enter
the estimate of relative kinematic probability of disk versus bulge
lenses.  See Equation~(\ref{eqn:pkineval}).
}
\label{fig:rgb}
\end{figure}

As we describe below, the values of $f^{(-,-)}_{\rm bulge}$ and $f^{(+,+)}_{\rm bulge}$
can be evaluated purely empirically by counting RC (or RGB) stars
in small areas in the neighborhoods of the two solutions and comparing
these values to the total sample.  However, the same principle cannot
be applied to find $f^{(-,-)}_{\rm disk}$ and $f^{(+,+)}_{\rm disk}$ by
counting MS stars.  This is because the MS stars come from many different
distances $D$ along the line of sight.  If, as in many Galactic models
used to carry out Bayesian analyses (e.g., \citealt{han95}), the rotation
curve is assumed flat, 
then the mean proper motion of disk stars at any distance
will always be the same.  For this reason, it is appropriate to
use the peak of the observed MS proper motions to evaluate the
mean proper motion of disk stars at the distance of the lens, $D_L$.
However, if (as also usually assumed) the velocity dispersions are 
independent of distance, then
the proper-motion dispersions of disk stars
scale $\sigma(\mu)\propto D^{-1}$.  Since disk stars that are closer
are systematically brighter at fixed luminosity (due both to proximity
and lower extinction), the sample of MS stars is highly biased toward
nearby stars with larger proper-motion dispersions that are quite
unrepresentative
of stars at $D_L\sim 7\,\kpc$.  It is for this reason that we evaluated
$c_{ij}$, including both intrinsic dispersion and measurement errors.
Therefore, we can write
\begin{equation}
P^{\rm kin}_{\rm disk} = f^{(-,-)}_{\rm disk} + Q f^{(+,+)}_{\rm disk}
= {\exp(-\chi^2_{\rm offset}/2)\over 2\pi \sqrt{|c|}} +
Q{\exp(-\chi^2_{{\rm offset},(+,+)}/2)\over 2\pi \sqrt{|c^{(+,+)}|}}
\rightarrow
{\exp(-\chi^2_{\rm offset}/2)\over 2\pi \sqrt{|c|}}
\label{eqn:pkindisk}
\end{equation}
where we have dropped the second term in the final expression
because $\chi^2_{{\rm offset},(+,+)}=55$.
Noting that $\pi|c|^{1/2}$ is just the area of the error ellipse, we can
now express the ratio of kinematic probabilities as
\begin{equation}
{P^{\rm kin}_{\rm disk}\over P^{\rm kin}_{\rm bulge}}
= {\exp(-\chi^2_{\rm offset}/2)/2\over 
(N_{\rm bulge}^{(-,-)} + Q N_{\rm bulge}^{(+,+)})/N_{\rm bulge}}\rightarrow 3.0\pm 0.3 \,,
\label{eqn:pkineval}
\end{equation}
where we have made the evaluation using the RGB sample with 
$N_{\rm bulge}=2123$, and where $N_{\rm bulge}^{(-,-)}=57$ and
$N_{\rm bulge}^{(+,+)}=57$ are the numbers of RGB stars in the two
ellipses shown in Figure~\ref{fig:rgb}.

Before continuing, we note that we performed a similar
test, but restricted to the 1008 RC stars, which are basically a subset
of the RGB sample, but even less prone to contamination from foreground
disk stars.  We found 23 and 36 stars in the $(-,-)$ and $(+,+)$
ellipses.  Inserting these numbers into Equation~(\ref{eqn:pkineval})
we obtain $P^{\rm kin}_{\rm disk}/P^{\rm kin}_{\rm bulge}=2.9\pm 0.4$., which
(even considering that these are overlapping samples) is consistent
at the $1\,\sigma$ level.  Given that the sign of the difference
is the opposite of what one would expect from greater contamination of the
RGB sample, we adopt the RGB value (i.e., Equation~(\ref{eqn:pkineval})).

A more detailed analysis would require a more precise Galactic model than
presently exists.  Below, we outline some of the issues that would have
to be addressed by such a model, but the key point is that vastly 
improved models are likely to be available within a year based on
the Gaia DR2 data release.  Hence, given the delicacy of the required
calculations, it is premature to carry them out based on current Galactic 
models.

Here, we just illustrate some of the issues that need to be considered.
The first issue is that the assumption of constant velocity dispersion
may well be incorrect.  The scale heights of edge-on disks 
of external galaxies appear to be constant as a function of radius,
while the radial density profiles are eponymously ``exponential''.
These simple observations argue for a vertical velocity dispersion
that scales roughly as the square root of surface density.  However,
by chance, any such adjustment would have a small effect in the present
case.  To see this, first note that (again by chance), 
$c_{ij}^{\rm pred} \simeq 2c_{ij}^{\rm meas}$.  Therefore, if we were to, say,
double the dispersions (i.e., multiply $c_{ij}^{\rm pred}$ by a factor four),
this would increase $c$ by a factor 3.0.  This would then
change $P_{\rm disk}^{\rm kin}$ by a factor: $\exp(\chi^2_{\rm offset}/3)/3=0.83$.

A second kinematic issue arises from possible streaming motions along
the bar, which might for example be responsible for the elongated
contours along the direction of the Galactic plane
in the RC distribution shown in Figure~\ref{fig:smoothpm}.  The lens
must be in front of the source (by $D_{LS}\sim 1\,\kpc$).  Hence, if
this streaming motion were primarily ``outward'' for stars in the
closer side of the bar, then there would be a relatively big population
of potential bulge lenses with proper motions strongly aligned with
Galactic rotation.  On the other hand, if the outward streaming motion
were mainly on the more distant side of the bar (and the nearer side
was streaming toward the Galactic center), then a bulge lens would
be much less likely.

Finally, the density distribution of both the bar and the disk
in this region must be estimated much more precisely than at present.
For example, a very narrow bar would make it difficult to accommodate
both a lens and source, with $D_{LS}\sim 1\,\kpc$.  Moreover, it is
possible that the disk in the immediate neighborhood of the bar is depleted
relative to an exponential profile, due to action by the bar.

For these reasons, we defer a detailed calculation of 
$P_{\rm disk}/P_{\rm bulge}$ until more precise models are developed
on the basis of the Gaia DR2 release.

\section{{A New Approach to Breaking the 2L1S/1L2S Degeneracy}
\label{sec:method}}

The space-based and ground-based light curves are each reasonably
well fit to 1L2S models.   These models have six non-linear parameters,
$[(t_0,u_0)_{1,2},t_\e,q_f]$.  Because there are two sources, there
are two pairs of $(t_0,u_0)$, one for each source.  The flux ratio $q_f$
is assumed to be the same for all observations in the same band (in our
case $I$ for ground-based data and $L$ for {\it Spitzer} data).  For fits
with more than one band, there is one ``$q_f$'' for each band.
Table~\ref{tab:1L2S} shows the fit parameters for {\it Spitzer}-only,
ground-only, and joint 1L2S fits.  

Comparing the $\chi^2$ values to those in Table~\ref{tab:free_rho},
we see that $\Delta\chi^2\equiv\chi^2({\rm 1L2S})-\chi^2({\rm 2L1S})$ 
takes on values
$\Delta\chi^2=(+55,+19,+804)$ for {\it Spitzer}-only, ground-only,
and {\it Spitzer}+ground data sets, respectively.  That is, whereas
the 2L1S and 1L2S geometries yield models with qualitatively
comparable $\chi^2$ values when the ground-based data are analyzed alone,
and are moderately-well distinguished based on {\it Spitzer} data alone,
the 1L2S solution is decisively excluded for the joint fit to all data.

As a first step toward  understanding the physical origin of this effect,
we note that whereas for 2L1S,
$\chi^2_{\rm joint,2L1S} = \chi^2_{spitzer, \rm2L1S} + \chi^2_{\rm ground,2L1S} -2$,
for 1L2S we find
$\chi^2_{\rm joint,1L2S} = \chi^2_{spitzer, \rm1L2S} + \chi^2_{\rm ground,1L2S} +725$.
The approximate equality,
$\chi^2_{\rm joint,2L1S} \simeq \chi^2_{spitzer, \rm2L1S} + \chi^2_{\rm ground,2L1S}$,
is expected from
the fact (already noted in Section~\ref{sec:rhozero})
that the {\it Spitzer}-only and ground-only 2L1S solutions are compatible
with each other.  This leads us to investigate whether the analogous
1L2S solutions are incompatible with each other.

\begin{figure}
\epsscale{0.90}
\plotone{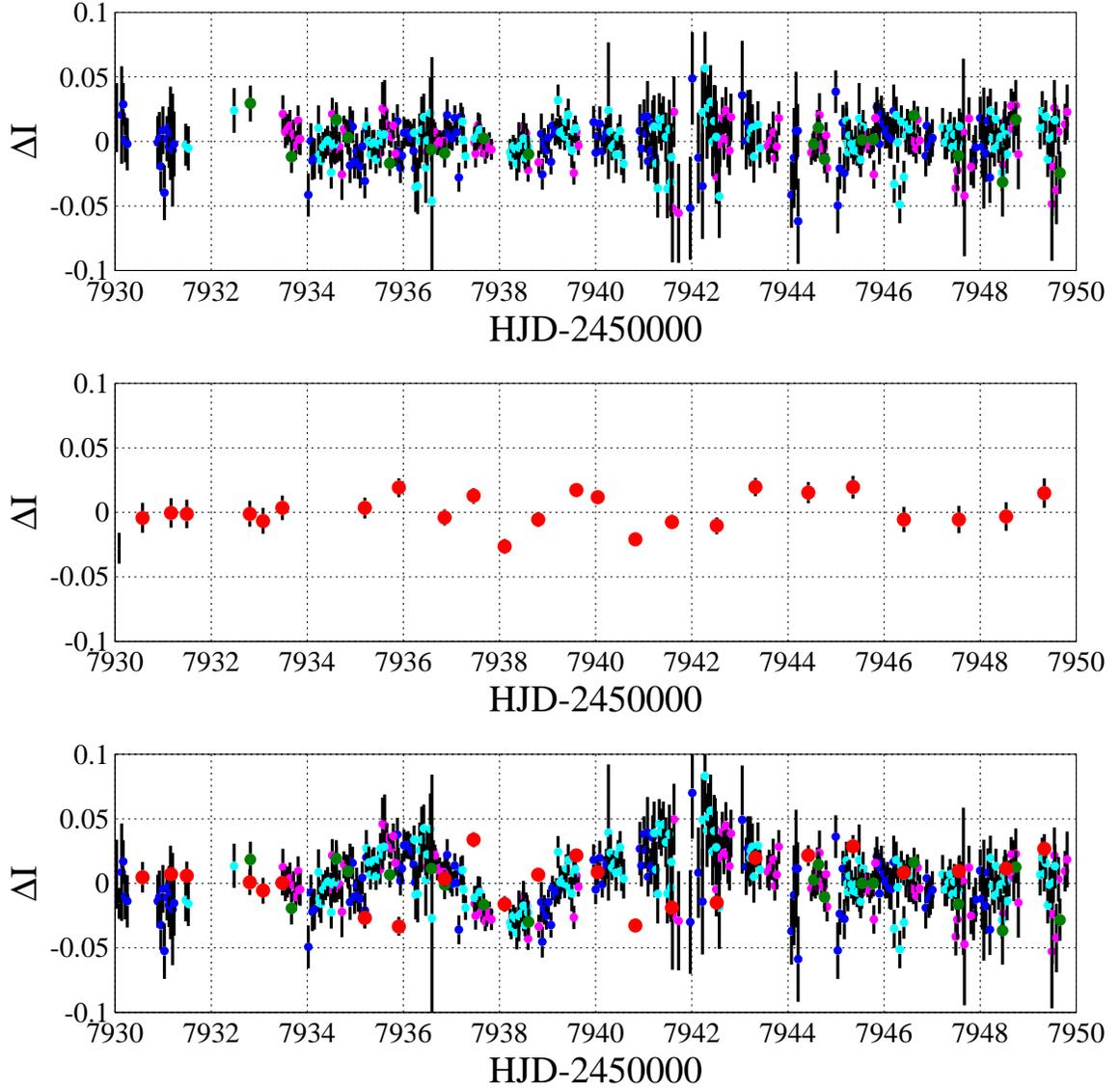}
\caption{Residuals to single-lens/binary source (1L2S) models
for three cases: ground-only, {\it Spitzer}-only, and joint fits
to all of the data.  While the residuals shown in the upper two panels
are somewhat worse than those shown for the 2L1S case in 
Figure~\ref{fig:lc}, the residuals for the joint fit (bottom panel)
are dramatically worse.  
This is because the separate solutions are consistent with each other
for 2L1S, but not for 1L2S. See Section~\ref{sec:method}.
}
\label{fig:1L2S}
\end{figure}

To pursue this question further, we introduce for 1L2S models
the vector offset within the Einstein ring of the two sources,
\begin{equation}
(\Delta\tau,\Delta\beta)_{\rm 1L2S}\equiv 
\biggl({t_{0,2}-t_{0,1}\over t_\e},u_{0,2}-u_{0,1}\biggr)\,.
\label{eqn:dtaubeta}
\end{equation}
Ignoring the very small motion of the binary source during the
few days between the passage of the lens by the sources, these vector
offsets should be the same as seen by two different observers.  However,
we find from Table~\ref{tab:1L2S},
\begin{equation}
(\Delta\tau,\Delta\beta)_{\rm ground,1L2S}=(+0.31,+0.15);
\qquad
(\Delta\tau,\Delta\beta)_{spitzer,\rm 1L2S}=(+0.19,+0.09)\,.
\label{eqn:dtbeval}
\end{equation}
In particular, we note that the offsets in $t_0$ differ by about
1.7 days between models of the two data sets, whereas the
errors in the individual measured values are all less than 0.04 days. 
Hence, in the
joint solution, the two separately-successful 1L2S models cannot
be accommodated with a single $(\Delta\tau,\Delta\beta)_{\rm 1L2S}$.
This inconsistency is illustrated by the residuals to the three
fits, which are shown in Figure~\ref{fig:1L2S}.

The fundamental origin for this incompatibility is that the magnification
(actually, logarithm of magnification) falls off at different
rates for binary-lens (or multi-lens) cusps than it does for
point lenses.  Of course, it is possible to arrange special geometries
that avoid this problem.  For example, if the impact parameter is
the same as seen by the two observatories, so that the same
event essentially repeats at a later time, which can occasionally happen
\citep{ob140124}, then any 1L2S/2L1S degeneracy
(or indeed any other degeneracy) will persist.  However, in the more
generic case, we should expect that this degeneracy can be broken
provided that both observatories have some sensitivity to both bumps.

\section{{Discussion}
\label{sec:discuss}}

OGLE-2017-BLG-1140 is the first anomalous microlensing event for which
observation of the anomaly from both Earth and {\it Spitzer} was essential
to the proper characterization of the anomaly.  In particular, we showed
that only by combining both data sets was it possible to decisively
discriminate between the 2L1S and 1L2S interpretations.  If this
indeed represents a new path toward breaking this degeneracy,
why is it appearing here for the first time?

For randomly selected microlensing events observed from two platforms, 
the relative strength of the anomalies observable at the two sites
should be likewise randomly distributed.  However,
among the 18 published 2L1S events observed by {\it Spitzer},
OGLE-2017-BLG-1140 is only the second one for which
the anomaly was stronger as observed by {\it Spitzer} than
from the ground. In the other case, OGLE-2017-BLG-1130 (which
coincidentally was alerted by OGLE and chosen as a ``secret''
{\it Spitzer} target at exactly the same times as OGLE-2017-BLG-1140),
the anomaly was seen from {\it Spitzer} only\footnote{For two other
events, the anomaly was of comparable strength as seen from {\it Spitzer}
and the ground: OGLE-2014-BLG-0124 \citep{ob140124} and OGLE-2015-BLG-1285
\citep{ob151285}.  Moreover, for two 1L1S events, finite-source effects
were observed by {\it Spitzer} but not from the ground: 
OGLE-2015-BLG-0763 \citep{ob150763} and
OGLE-2015-BLG-1482 \citep{ob151482}.} \citep{ob171130}.  

There are four factors that explain this apparent discrepancy.
First, of the 18 2L1S events, five had short timescale anomalies
due to a planet\footnote{
OGLE-2015-BLG-0966 \citep{ob150966},
OGLE-2016-BLG-1067 \citep{ob161067},
OGLE-2016-BLG-1190 \citep{ob161190},
OGLE-2016-BLG-1195 \citep{ob161195a,ob161195b}, and
OGLE-2017-BLG-1140 (this work).}.  Because {\it Spitzer}'s cadence
has typically been $\Gamma\sim 1\,{\rm day}^{-1}$, it cannot in general
be expected to characterize short-term anomalies in the absence
of dense ground-based data over the anomaly.  That said, it should
be pointed out that OGLE-2017-BLG-1140 is one of these five events.

Second, the majority of {\it Spitzer} targets are near peak
or have already peaked as seen from the ground at the time of
the onset of {\it Spitzer} observations.  This alone would
imply that half or more of the anomalies that would be visible
from {\it Spitzer}'s location are in fact missed by {\it Spitzer}
observations.  This late onset follows from the delay in 
{\it Spitzer} uploads (see Figure~1 of \citealt{ob140124})
and the difficulty of recognizing and reliably choosing microlensing
events based on their early evolution.

Third, due to the direction of Galactic rotation expressed in 
equatorial coordinates, more disk lenses are traveling east than
west, meaning that they peak later as seen from {\it Spitzer},
which lies to the west of Earth.  In itself, this is a relatively
minor effect, but it exacerbates the previous one.

Fourth, {\it Spitzer} can observe targets that are near the ecliptic for
a maximum of 38 days.  Hence, for long events, anomalies can take
place outside of the {\it Spitzer} window.

Taken together, these four effects mean that the new channel for
resolving the 2L1S/1L2S degeneracy will not appear on a routine
basis in {\it Spitzer} microlensing events.  Nevertheless, it
is worth noting that despite the relatively weak appearance of the
OGLE-2017-BLG-1140 anomaly in ground-based data, the addition of the also fairly
modest signal from the {\it Spitzer} anomaly dramatically improved the
confidence of the result.  Further, although the anomaly was
recognized in ground-based data soon after it occurred, the
event was not systematically analyzed because it appeared
to have insurmountable degeneracies.  Therefore, it is quite
possible that other archival events with even weaker, less noticeable,
anomalies can also yield interesting, unexpected results.
Moreover, this same principle can be applied to future parallax-satellite
missions, including {\it WFIRST} \citep{spergel13} as well as other
missions that are yet unplanned.

\section{{Conclusions}
\label{sec:end}}

We have presented OGLE-2017-BLG-1140Lb, 
a microlensing extrasolar planet detected 
combining ground-based survey, OGLE and KMTNet,
and space-based, \emph{Spitzer}, data.
From the modeling point of view
this event is of particular interest. For the first time
\emph{Spitzer}, besides providing the measure
of the microlensing parallax, is essential for the characterization
of the planetary system. Indeed,
a deviation from the single-lens single-source (1L1S)
Paczy\'nski shape is apparent both from ground
(specifically, KMTNet), 
and space-based data which however, separately,
are each reasonably well fit by either a single-lens
binary-source (1L2S) or a binary-lens
single-source (2L1S), planetary, model.
The analysis then leads us to show how
the microlensing parallax opens a new path for breaking
this classic degeneracy \citep{gaudi98} when
combining ground and space-based data.
Specifically, 
we find that that the 1L2S solution is ruled out by $\Delta\chi^2\sim 800$
by the combined space/ground analysis, which is a factor 10 higher than
the $\Delta\chi^2\sim 80$ from the sum of the ground and space analyses
considered separately.
As for the 2L1S planetary model, the system
can be indipendently characterized 
by \emph{Spitzer}, whose trajectory
passes closer to the caustic structure,
and ground-based data, leading
roughly to the same configuration
(except that ground-based data alone do not 
allow to constrain the finite source effect
parameter, $\rho$). As we show, however,
the combination of the two data sets
puts a stronger constraint on the caustic
structure (the binary parameter $(s,q)$, i.e.,
the projected separation of the two
lenses and their mass ratio, resulting specifically
in a ``resonant'' configuration) and indirectly
also on $\rho$. The measurement of $\rho$,
together with the photometric characterization
of the source, and the measurement of the
microlens parallax allow us to determine
the physical parameters of the system.
Specifically we find $M_{\rm host} = 0.21\pm 0.03\,M_\odot$
and $M_{\rm planet} = 1.6^{+0.4}_{-0.3}\,M_{\rm jup }$, 
for a lens-to-source distance $D_{LS}\simeq 1\,$kpc, and
a planet-host separation
$a_\perp = 1.02\pm 0.15\,\au$, well beyond the system snow line.
We show that the lens proper motion analysis is consistent with the
lens lying in the Galactic disk, although a Bulge lens is not
ruled out.
In the framework of the \emph{Spitzer} microlensing
survey, OGLE-2017-BLG-1140Lb is the fifth planet to enter
the sample for the determination of the Galactic
distribution of exoplanets \citep{21event,zhu17}.

The discovery of OGLE-2017-BLG-1140Lb, a super-Jupiter mass planet
orbiting a M-dwarf (beyond the system snow line), is also relevant
in the larger framework of the microlensing statistical 
census of exoplanets (e.g., \citealt{gaudi12,gould16}). 
Indeed, out of 58 (microlensing) planets
currently known\footnote{https://exoplanetarchive.ipac.caltech.edu.}, 
11 belong to that same
class (specifically for a host mass $0.08<M/M_\odot<0.5$,
and a planet mass larger than that of Jupiter, e.g., \citealt{yossi14}).
OGLE-2017-BLG-1140Lb adds to this sample,
and it is the fifth one belonging to the subsample
of those with 
microlens-parallax-based mass measurements, which are substantially
more accurate.
Notwithstanding the microlensing observational bias
for the detection of such planetary systems (e.g., \citealt{batista11}),
because of the abudance of M-dwarf and 
of the detection efficiency's increase with $q$,
the abundance of these systems, about 20\% of all 
microlensing planets, remains a challenge
for current planet formation theories (e.g., \citealt{lissauer10}).

\acknowledgments 
The OGLE project has received funding from the National Science Centre,
Poland, grant MAESTRO 2014/14/A/ST9/00121 to AU.
Work by YKJ, and AG were supported by AST-1516842 from the US NSF.
IGS, and AG were supported by JPL grant 1500811.
This research has made use of the KMTNet system operated by the Korea
Astronomy and Space Science Institute (KASI) and the data were obtained at
three host sites of CTIO in Chile, SAAO in South Africa, and SSO in
Australia.
Work by YS was supported by an appointment to the NASA Postdoctoral
Program at the Jet Propulsion Laboratory, California Institute of
Technology, administered by Universities Space Research Association
through a contract with NASA.
 Work by C. Han was supported by the grant
(2017R1A4A1015178) of National Research Foundation of Korea.
This work is based (in part) on observations made with the Spitzer Space
Telescope, which is operated by the Jet Propulsion Laboratory, California
Institute of Technology under a contract with NASA. Support for this work
was provided by NASA through an award issued by JPL/Caltech.
This work was partially supported by NASA contract NNG16PJ32C.

----

\begin{deluxetable}{lrrrr}

\tablecaption{2L1S Solutions with $\rho=0$} 

\tablehead{Parameter & \multicolumn{1}{c}{\emph{Spitzer}} & \multicolumn{1}{c}{Ground} & \multicolumn{2}{c}{\emph{Spitzer} and ground}}
\startdata
$\chi^2$/dof & 44.8/35 & 2964.6/2936 &  3024.9/2975 & 3025.9/2975 \\ [5pt]
 & &  &  $(+,+)$ & $(-,-)$ \\ [5pt]
\hline
$t_0$ [HJD-2457940.] & $-0.201_{-0.011}^{+0.011}$  & $0.7645_{-0.0066}^{+0.0063}$  & $0.7604_{-0.0062}^{+0.0060}$ & $0.7606_{-0.0062}^{+0.0061}$ \\  [5pt]
$u_0$ & $0.1343_{-0.0082}^{+0.0083}$  & $0.2373_{-0.0055}^{+0.0044}$  &  $0.2364_{-0.0042}^{+0.0040}$  &  $-0.2365_{-0.0040}^{+0.0042}$ \\ [5pt]
$t_\mathrm{E}$ [days] & $14.58_{-0.63}^{+0.68}$  & $14.74_{-0.13}^{+0.17}$  & $14.74_{-0.13}^{+0.14}$ & $14.74_{-0.13}^{+0.14}$ \\ [5pt]
$\rho$  & 0  & 0  & 0 & 0 \\ 
$\pi_\mathrm{E,N}$ & - & - & $-0.0793_{-0.0016}^{+0.0016}$ & $0.0799_{-0.0016}^{+0.0016}$\\ [5pt]
$\pi_\mathrm{E,E}$ & - & - & $ 0.0524_{-0.0007}^{+0.0007}$ & $ 0.0520_{-0.0007}^{+0.0007}$\\ [5pt]
$\alpha$ [rad] & $2.561_{-0.013}^{+0.013}$  & $2.545_{-0.011}^{+0.012}$  &$2.5463_{-0.0083}^{+0.0085}$ &$-2.5462_{-0.0083}^{+0.0082}$ \\ [5pt]
$s$ & $0.815_{-0.021}^{+0.019}$ &  $0.855_{-0.019}^{+0.017}$  &$0.831_{-0.012}^{+0.012}$ & $0.831_{-0.012}^{+0.012}$ \\ [5pt]
$q$ & $0.0099_{-0.0013}^{+0.0015}$ &  $0.0079_{-0.0010}^{+0.0012}$ &  $0.0090_{-0.0008}^{+0.0009}$ &  $0.0090_{-0.0008}^{+0.0009}$\\ [5pt]
\hline
$f_\mathrm{s,OGLE}$ & - & $1.131\pm 0.024$& $1.127 \pm 0.020$ & $1.127 \pm 0.020$\\ [5pt]
$f_\mathrm{b,OGLE}$ & - & $0.237\pm 0.024$& $0.241 \pm 0.020$ & $0.241 \pm 0.020$\\ [5pt]
$f_{\mathrm{s}, Spitzer}$ & $17.4 \pm 1.1$ & -& $17.4\pm 0.3$& $17.4\pm 0.3$\\ [5pt]
$f_{\mathrm{b}, Spitzer}$ & $2.4\pm 0.9$& -& $2.4\pm 0.3$& $2.5\pm 0.3$\\ [5pt]
$I-L$ & - & -  & $2.971 \pm 0.007$ & $2.971\pm 0.007$\\ [5pt]
\enddata
\label{tab:fixed_rho}
\end{deluxetable}

\begin{deluxetable}{lrrrr}

\tablecaption{2L1S Solutions with $\rho\ne 0$} 

\tablehead{Parameter & \multicolumn{1}{c}{\emph{Spitzer}} & \multicolumn{1}{c}{Ground} & \multicolumn{2}{c}{\emph{Spitzer} and ground}}
\startdata
$\chi^2$/dof & 39.4/34 & 2964.0/2935 &  3002.0/2974 & 3001.0/2974\\ [5pt]
 & &  &  $(+,+)$ & $(-,-)$ \\ [5pt]
\hline
$t_0$ [HJD-2457940.] & $-0.195_{-0.010}^{+0.010}$  & $0.7677_{-0.0066}^{+0.0065}$  & $0.7698_{-0.0059}^{+0.0059}$ & $0.7701_{-0.0058}^{+0.0059}$ \\  [5pt]
$u_0$ & $0.1390_{-0.0082}^{+0.0085}$  & $0.2395_{-0.0055}^{+0.0040}$  &  $0.2389_{-0.0040}^{+0.0034}$ &  $-0.2392_{-0.0032}^{+0.0038}$ \\ [5pt]
$t_\mathrm{E}$ [days] & $14.38_{-0.61}^{+0.65}$  & $14.68_{-0.11}^{+0.17}$  & $14.70_{-0.11}^{+0.14}$ & $14.69_{-0.11}^{+0.13}$ \\ [5pt]
$\rho$ & $0.0241_{-0.0078}^{+0.0058}$ &  - &  $0.0269_{-0.0034}^{+0.0026}$ &  $0.0270_{-0.0027}^{+0.0024}$ \\
$\pi_\mathrm{E,N}$ & - & - & $-0.0782_{-0.0015}^{+0.0016}$ & $0.0789_{-0.0015}^{+0.0014}$\\ [5pt]
$\pi_\mathrm{E,E}$ & - & - & $ 0.0531_{-0.0008}^{+0.0007}$ & $ 0.0528_{-0.0007}^{+0.0007}$\\ [5pt]
$\alpha$ [rad] & $2.557_{-0.012}^{+0.011}$  & $2.540_{-0.011}^{+0.012}$  &$2.539_{-0.0074}^{+0.0076}$ &$-2.5388_{-0.0069}^{+0.0067}$ \\ [5pt]
$s$ & $0.857_{-0.028}^{+0.027}$ &  $0.871_{-0.022}^{+0.026}$  &$0.870_{-0.014}^{+0.014}$ &$0.871_{-0.013}^{+0.012}$ \\ [5pt]
$q$ & $0.0080_{-0.0012}^{+0.0014}$ &  $0.0072_{-0.0011}^{+0.0012}$ &  $0.0073_{-0.0007}^{+0.0008}$ &  $0.0072_{-0.0006}^{+0.0007}$ \\ [5pt]
\hline
$f_\mathrm{s,OGLE}$ & - & $1.138\pm 0.024$& $1.136 \pm 0.019$ & $1.138 \pm 0.018$\\ [5pt]
$f_\mathrm{b,OGLE}$ & - & $0.231\pm 0.024$& $0.232 \pm 0.019$ & $0.231 \pm 0.018$\\ [5pt]
$f_{\mathrm{s}, Spitzer}$ & $17.9 \pm 1.1$ & -& $17.7\pm 0.3$& $17.7\pm 0.3$\\ [5pt]
$f_{\mathrm{b}, Spitzer}$ & $1.9\pm 1.0$& -& $2.0\pm 0.3$& $2.0\pm 0.3$\\ [5pt]
$I-L$ & - & -  & $2.982 \pm 0.008$ & $2.980\pm 0.007$\\ [5pt]
\enddata
\label{tab:free_rho}
\end{deluxetable}

\begin{deluxetable}{lrr}
\tablewidth{0pt}
\tablecaption{Physical parameters}
\tablehead{Parameter & \multicolumn{1}{c}{$(+,+)$} & \multicolumn{1}{c}{$(-,-)$}}
\startdata
\hline
$M_\mathrm{host}$ ($\mathrm{M}_\odot$) &$0.213_{-0.027}^{+0.036}$&$0.211_{-0.025}^{+0.032}$\\[5pt]
$M_\mathrm{planet}$ ($\mathrm{M}_\mathrm{Jup}$)  &$1.62_{-0.29}^{+0.41}$&$1.59_{-0.26}^{+0.35}$\\[5pt]
$D_{8.3}$ (kpc) & $7.36_{-0.14}^{+0.11}$&$7.36_{-0.12}^{+0.10}$\\[5pt]
$\theta_\mathrm{E}$ (mas) & $0.164_{-0.020}^{+0.028}$ & $0.163_{-0.019}^{+0.024}$\\ [5pt]
$\pi_\mathrm{E}$ & $0.0946_{-0.0016}^{+0.0014}$ & $0.0949_{-0.0015}^{+0.0014}$ \\[5pt]
$\pi_\mathrm{rel}$ (mas) & $0.0155_{-0.0019}^{+0.0027}$&  $0.0154_{-0.0019}^{+0.0023}$\\ [5pt]
$\mu_\mathrm{rel}$ (mas yr$^{-1}$) & $4.07_{-0.50}^{+0.69}$& $4.04_{-0.48}^{+0.60}$\\ [5pt]
${\tilde v}_\mathrm{hel,N}$ (km s$^{-1}$) & $-1030.7_{-8.7}^{+8.4}$&$1030.9_{-8.2}^{+8.3}$\\ [5pt]
${\tilde v}_\mathrm{hel,E}$ (km s$^{-1}$) & $728_{-13}^{+13}$& $719_{-13}^{+13}$ \\ [5pt]
\enddata
\label{tab:phys_par}
\end{deluxetable}

\begin{deluxetable}{lrrrr}
\tabletypesize{\small}
\tablecaption{1L2S Solutions} 

\tablehead{Parameter & \multicolumn{1}{c}{\emph{Spitzer}} & \multicolumn{1}{c}{Ground} & \multicolumn{2}{c}{\emph{Spitzer} and ground}}
\startdata
$\chi^2$/dof & 94.2/35 & 2985.3/2936 &  3834.8/2975 & 3804.9/2975\\ [5pt]
 & &  &  $(+,+)$ & $(+,-)$ \\ [5pt]
\hline
$t_\mathrm{E}$ [days] & $16.06_{-0.74}^{+0.81}$  & $15.12_{-0.15}^{+0.15}$  & $13.96_{-0.14}^{+0.14}$ & $14.49_{-0.17}^{+0.17}$ \\ [5pt]
$t_{0,1}$ [HJD-2457900.] & $37.022_{-0.023}^{+0.023}$  & $36.337_{-0.038}^{+0.040}$  & $37.698_{-0.031}^{+0.031}$ & $38.163_{-0.058}^{+0.057}$ \\  [5pt]
$t_{0,2}$ [HJD-2457900.] & $40.064_{-0.016}^{+0.017}$  & $41.026_{-0.019}^{+0.019}$  & $40.979_{-0.025}^{+0.025}$ & $41.361_{-0.047}^{+0.046}$ \\  [5pt]
$u_{0,1}$ & $0.0299_{-0.0027}^{+0.0027}$  & $0.0847_{-0.0048}^{+0.0050}$  &  $0.1808_{-0.0037}^{+0.0039}$ &  $0.2810_{-0.0081}^{+0.0088}$ \\ [5pt]
$u_{0,2}$ & $0.1237_{-0.0073}^{+0.0074}$  & $0.2309_{-0.0041}^{+0.0041}$  &  $0.2752_{-0.0046}^{+0.0048}$ &  $0.2282_{-0.0054}^{+0.0057}$ \\ [5pt]
$q_{f,I}$ &-&  $29.5_{-3.0}^{+3.3}$ & $15.6_{-1.5}^{+1.7}$& $2.75_{-0.20}^{+0.23}$\\ [5pt]
$q_{f,L}$ & $19.7_{-1.8}^{+2.1}$& - & $12.5_{-1.0}^{+1.0}$& $8.62_{-0.61}^{+0.70}$\\ [5pt]
$\pi_\mathrm{E,N}$ & - & - & $-0.1036_{-0.0023}^{+0.0021}$ & $-0.2795_{-0.0061}^{+0.0058}$\\ [5pt]
$\pi_\mathrm{E,E}$ & - & - & $ 0.0500_{-0.0012}^{+0.0012}$ & $0.0600_{-0.0021}^{+0.0021}$\\ [5pt]
\hline
$\Delta\tau_{\mathrm{Ground}}$ & - & $0.3099_{-0.0038}^{+0.0038}$ & $0.2355_{-0.0025}^{+0.0025}$ & $0.2209_{-0.0034}^{+0.0034}$\\[5pt]
$\Delta\tau_{Spitzer}$ & $0.1894_{-0.0092}^{+0.0092}$ & - & $0.2342_{-0.0026}^{+0.0026}$ & $0.2138_{-0.0031}^{+0.0031}$\\[5pt]
${\Delta}u_{0,\mathrm{Ground}}$ & - & $0.1462_{-0.0072}^{+0.0072}$ & $0.0944_{-0.0044}^{+0.0044}$ & $-0.531_{-0.0043}^{+0.0043}$\\[5pt]
${\Delta}u_{0,Spitzer}$ & $0.0939_{-0.0067}^{+0.0067}$ & - & $0.0892_{-0.0044}^{+0.0044}$ & $0.0666_{-0.0045}^{+0.0045}$\\[5pt]
$t_{0,1,Spitzer}$ [HJD-2457900.] & -  & -  & $36.865_{-0.030}^{+0.030}$ & $37.075_{-0.028}^{+0.028}$ \\  [5pt]
$t_{0,2,Spitzer}$ [HJD-2457900.] & -  & -  & $40.136_{-0.017}^{+0.017}$ & $40.172_{-0.017}^{+0.017}$ \\  [5pt]
$u_{0,1,Spitzer}$ & - & - &  $0.0523_{-0.0028}^{+0.0028}$ & $0.0663_{-0.0032}^{+0.0032}$  \\ [5pt]
$u_{0,2,Spitzer}$ & - & - &  $0.1415_{-0.0025}^{+0.0025}$ & $0.1329_{-0.0025}^{+0.0025}$  \\ [5pt]
\hline
$f_\mathrm{s,OGLE}$ & - & $1.069\pm 0.021$& $1.286 \pm 0.026$ & $1.172 \pm 0.029$\\ [5pt]
$f_\mathrm{b,OGLE}$ & - & $0.299\pm 0.021$& $0.082 \pm 0.026$ & $0.195 \pm 0.029$\\ [5pt]
$f_{\mathrm{s}, Spitzer}$ & $15.3 \pm 1.0$ & -& $18.0\pm 0.3$& $17.0\pm 0.3$\\ [5pt]
$f_{\mathrm{b}, Spitzer}$ & $4.2\pm 0.7$& -& $2.3\pm 0.3$& $3.2\pm 0.3$\\ [5pt]
\enddata
\label{tab:1L2S}
\end{deluxetable}


\begin{thebibliography}{99}

\bibitem[Alard \& Lupton(1998)]{alard98} Alard, C. \& Lupton, R.H.,1998, \apj, 503, 325

\bibitem[Albrow et~al.(2002)]{mb9947}Albrow, M.D., An, J., Beaulieu, J.-P., et~al.\ 2002, \apj, 572, 1031

\bibitem[Afonso et~al.(2000)]{ms9801}Afonso, C., Alard, C., Albert, J.N., et~al. 2000, \apj, 532, 340

\bibitem[Albrow et~al.(2009)]{albrow09}Albrow, M.\ D., Horne, K., Bramich, D.\ M., et~al.\ 2009, \mnras, 397, 2099

\bibitem[An(2005)]{an05}An, J.H., \mnras, 356, 1409

\bibitem[Batista et~al.(2011)]{batista11}Batista, V., Gould, A., Dieters, S. et~al. 2011, \aap, 529, A102

\bibitem[Beaulieu et~al.(2006)] {ob05390}Beaulieu, J.-P. Bennett, D.P., Fouqu\'e, P. et~al. 2006, Nature, 439, 437

\bibitem[Bensby et~al.(2013)]{bensby13} Bensby, T. Yee, J.C., Feltzing, S.\ et~al.\ 2013, \aap, 549, A147

\bibitem[Bessell \& Brett(1988)]{bb88} Bessell, M.S., \& Brett, J.M.\ 1988, \pasp, 100, 1134

\bibitem[Bond et~al.(2017)]{ob161195a} Bond, I.A., Bennett, D.P., Sumi, T. et~al.\ 2017, \mnras, 469, 2434

\bibitem[Bozza(2010)]{bozza10}Bozza, V., 2010, \mnras, 408, 2188

\bibitem[Calchi Novati et~al.(2015a)]{21event}  Calchi Novati, S., Gould, A., Udalski, A., et~al., 2015a, \apj, 804, 20

\bibitem[Calchi Novati et~al.(2015b)]{170event} Calchi Novati, S., Gould, A., Yee, J.C., et~al. 2015b, \apj, 814, 92

\bibitem[Calchi Novati et~al.(2018)]{ob161067} Calchi Novati, S., Suzuki, D., Udalski, A., et~al. 2018, submitted, arXiv:1801.05806

\bibitem[Chung et~al.(2017)]{ob151482} Chung, S.-J., Zhu, W., Udalski, A., 2017, \apj, 838, 154

\bibitem[D'Angelo et~al.(2010)]{lissauer10} D'Angelo, G., Durisen, R.~H,
\& Lissauer, J.~J. 2010, {Giant Planet Formation}, ed. S.~{Seager} 319

\bibitem[Dominik(1999)]{dominik99} Dominik, M. 1999, \aap, 349, 108

\bibitem[Dong et~al. (2009)]{subo09}Dong, S., Gould, A., Udalski, A., 
et~al. 2009, \apj, 695, 970

\bibitem[Gaudi(1998)]{gaudi98} Gaudi, B.~S.\ 1998, \apj, 506, 533

\bibitem[Gaudi(2012)]{gaudi12} Gaudi, B.~S.\ 2012, \araa, 50, 411

\bibitem[Gould(1992a)]{gould92a} Gould, A. 1992a, \apjl, 386, 5

\bibitem[Gould(1992b)]{gould92} Gould, A. 1992b, \apj, 392, 442

\bibitem[Gould(1994)]{gould94} Gould, A. 1994, \apjl, 421, L75

\bibitem[Gould(1997)]{gould97} Gould, A. 1997, The Hollywood Strategy for Microlensing Detection of Planets, in Variables Stars and the Astrophysical Returns of the Microlensing Surveys. Eds. R. Ferlet, J.-P. Maillard and B. Raban. Gif-sur-Yvette, France : Editions Frontieres, p.125

\bibitem[Gould(2000)]{gould00} Gould, A. 2000, \apj, 542, 785

\bibitem[Gould(2004)]{gould04} Gould, A. 2004, \apjl, 606, 319

\bibitem[Gould(2016)]{gould16} Gould, A. 2016, in Astrophysics 
and Space Science Library, Vol. 428, Methods
  of Detecting Exoplanets: 1st Advanced School on Exoplanetary Science, ed.
  V.~{Bozza}, L.~{Mancini}, \& A.~{Sozzetti}, 135

\bibitem[Gould et~al.(2013)]{prop2013} Gould, A., Carey, S., \& Yee, J. 2013, 2013spitz.prop.10036

\bibitem[Gould et~al.(2014)]{prop2014} Gould, A., Carey, S., \& Yee, J. 2014, 2014spitz.prop.11006

\bibitem[Gould et~al.(2015a)]{prop2015a} Gould, A., Yee, J., \& Carey, S., 2015a, 2015spitz.prop.12013

\bibitem[Gould et~al.(2015b)]{prop2015b} Gould, A., Yee, J., \& Carey, S., 2015b, 2015spitz.prop.12015

\bibitem[Gould et~al.(2016)]{prop2016} Gould, A., Yee, J., \& Carey, S., 2016, 2015spitz.prop.13005

\bibitem[Griest \& Safizadeh(1998)]{griest98} Griest, K.\ \& Safizadeh, N.\ 1998, \apj, 500, 37

\bibitem[Han \& Gould(1995)]{han95} Han, C. \& Gould, A.\ 1995, \apj, 447, 53

\bibitem[Han et~al.(2018)]{ob170482}Han, C., Sumi, T., Udalski, A., et~al. 2018, submitted

\bibitem[Hwang et~al.(2017)]{ob151459} Hwang, K.-H., Udalski, A., Bond, I.A., et~al. 2017, submitted arXiv:1711.09651

\bibitem[Hwang et~al.(2018)]{ob170173} Hwang, K.-H., Udalski, A., Shvartzvald, Y. et~al. 2018, \aj, 155, 20 

\bibitem[James \& Roos (1975)]{minuit75} James, F., \& Roos, M.\ 1975, 
CoPhC, 10, 343

\bibitem[Jung et~al.(2017)]{ob160733}Jung, Y.\ K., Udalski, A., Yee, J.C., et~al.\ 2017 \aj, 153, 129

\bibitem[Kervella et~al.(2004)]{kervella04} Kervella, P., Bersier, D., Mourard, D. et~al \ 2004, \aap, 428, 587


\bibitem[Kim et~al.(2016)]{kmtnet} Kim, S.-L., Lee, C.-U., Park, B.-G., et~al.  2016, JKAS, 49, 37

\bibitem[Nataf et~al.(2013)]{nataf13} Nataf, D.M., Gould, A., Fouqu\'e, P. et~al. 2013, \apj, 769, 88

\bibitem[Paczy\'nski(1986)]{pac86} Paczy\'nski, B.\ 1986, \apj, 304, 1

\bibitem[Refsdal(1966)]{refsdal66} Refsdal, S. 1966, \mnras, 134, 315

\bibitem[Ryu et~al.(2017)]{ob160693} Ryu, Y.-H., Udalski, A., Yee, J.C. et~al. 2017, \aj, 154, 247 

\bibitem[Ryu et~al.(2017)]{ob161190} Ryu, Y.-H., Udalski, A., Bond, I.A. et~al. 2018, \aj, 155, 40

\bibitem[Shvartzvald et~al.(2014)]{yossi14} Shvartzvald, Y., Maoz, D., Kaspi, S. et~al. 2014, \mnras, 439, 604

\bibitem[Shvartzvald et~al.(2015)]{ob151285} Shvartzvald, Y., Udalski, A., Gould, A.\ et~al.\ 2015, \apj, 814, 111

\bibitem[Shvartzvald et~al.(2017)]{ob161195b} Shvartzvald, Y., Yee, J.C., Calchi Novati, S.\ et~al.\ 2017, \apjl, 840, L3

\bibitem[Skowron et~al.(2011)]{ob09020}Skowron, J., Udalski, A., Gould, A. et~al.\ 2011, \apj, 738, 87

\bibitem[Skowron et~al.(2018)]{ob170373}Skowron, J., Ryu, Y.-H., Hwang, K.-H. et~al.\ 2018, Acta Astron. 68, 43


\bibitem[Spergel et~al.(2013)]{spergel13} Spergel, D.N., Gehrels, N., Breckinridge, J., et~al. 2013, arXiv:1305.5422

\bibitem[Street et~al.(2016)]{ob150966} Street, R., Udalski, A., Calchi Novati, S.\ et~al.\ 2016, \apj, 829, 93.

\bibitem[Tomaney \& Crotts (1996)]{tomaney96} Tomaney, A.~B. and Crotts, A.~P.~S. 1996, \aj, 112, 2872

\bibitem[Udalski(2003)]{ews2} Udalski, A. 2003, Acta Astron., 53, 291

\bibitem[Udalski et~al.(1994)]{ews1} Udalski, A.,Szymanski, M., Kaluzny, J., Kubiak, M., Mateo, M.,  Krzeminski, W., \& Paczy\'nski, B. 1994, Acta Astron., 44, 227

\bibitem[Udalski et~al.(2015a)]{ob140124} Udalski, A., Yee, J.C., Gould, A., et~al. 2015, \apj, 799, 237

\bibitem[Udalski et~al.(2015b)]{ogle-iv}Udalski, A., Szyma{\'n}ski, M.K., \& Szyma{\'n}ski, G. 2015, Acta Astron., 65, 1.


\bibitem[Udalski et~al.(2018)]{ob171434} Udalski, A.,Ryu, Y.-H., Sajadian, S., et~al.\ 2018, Acta Astron., 68, 1.

\bibitem[Wang et~al.(2018)]{ob171130}Wang, T., Calchi Novati, S., Udalski, A., et~al. 2018,  submitted, arXiv:1802.09023

\bibitem[Wo\'zniak(2000)]{wozniak2000} Wo\'zniak, P.~R. 2000, Acta Astron., 50, 421

\bibitem[Yee et~al.(2012)]{yee12} Yee, J.~C., Shvartzvald, Y., Gal-Yam, A., et~al. \ 2012, \apj, 755, 102

\bibitem[Yee et~al.(2015)]{yee15} Yee, J.C., Gould, A., Beichman, C., 2015, \apj, 810, 155

\bibitem[Yoo et~al.(2004)]{ob03262} Yoo, J., DePoy, D.~L., Gal-Yam, A.\ et~al.\ 2004, \apj, 603, 139

\bibitem[Zhu et~al.(2014)]{zhu14} Zhu, W., Penny, M., Mao, S., Gould, A., \& Gendron, R.\ 2014, \apj, 788, 73

\bibitem[Zhu et~al.(2015)]{ob141050} Zhu, W., Udalski, A., Gould, A., et~al. 2015, \apj, 805, 8

\bibitem[Zhu et~al.(2016)]{ob150763} Zhu, W., Calchi Novati, S., Gould, A., et~al. 2016, \apj, 825, 60

\bibitem[Zhu et~al.(2017)]{zhu17} Zhu, W., Udalski, A., Calchi Novati, S.,  et~al. 2017, \aj, 154, 210 

\end{thebibliography}
\end{document}